\documentstyle[12pt,epsfig]{article}
\makeatletter                    %allow @ in command names
\@addtoreset{equation}{section}  %reset equation count in each section
\makeatother                     %disallow @ in command names

\newskip\humongous \humongous=0pt plus 1000pt minus 1000pt

\newif\ifdtup

\def\be{\begin{equation}}
\def\ee{\end{equation}}

\begin{document}

\title{An Empirical Investigation of the Forward Interest Rate Term Structure}
\author{Andrew Matacz$^1$\thanks{Email: andrew.matacz@science-finance.fr} 
$~$and Jean-Philippe Bouchaud$^{1,2}$\thanks{Email: bouchaud@spec.saclay.cea.fr}\\
{\small $^1$ Science and Finance}\\
{\small 109-111 rue Victor Hugo} \\
{\small 92632 Levallois, France}\\
{\small http://www.science-finance.fr}\\
{\small $^2$ Service de Physique de l'Etat Condens\'e}\\
{\small CEA-Saclay, Orme des Merisiers} \\
{\small 91 191 Gif s/ Yvette, France}}
%\date{\small {\it Draft version -- Comments welcome --  July 1999}}
\maketitle

\begin{abstract}
In this paper we study empirically the Forward Rate Curve ({\sc frc}) of 5 different currencies.
We confirm and extend the findings of our previous investigation of the U.S. 
Forward Rate Curve. In particular, the average {\sc frc} follows a square-root law, with 
a prefactor related to the spot volatility, suggesting a Value-at-Risk like pricing. 
We find a striking correlation between the instantaneous {\sc frc} and the 
past spot trend over a certain time horizon, in agreement with the idea of an extrapolated trend effect.
We present a model which can be adequately calibrated to account for these effects.
\end{abstract}

%\newpage
\section{Introduction}

The statistical analysis of the time fluctuations of financial assets 
is a rapidly growing activity, partly due to the easy access of large
quantities of data, and also to the need of finding more adequate models
of the behaviour of financial markets. This is important both for risk control purposes, 
and for a better pricing and hedging of derivative products \cite{Hull,JPMP}. The case of the 
interest curve is particularly interesting and difficult, since one must 
find a consistent description of the time evolution of a one dimensional
{\it curve} (the interest rate for different maturities), rather than that of
a single point in the case of a stock or a currency. Interest rate derivatives
pricing and hedging, or Asset Liability Management, require an adequate model
for the evolution and deformation of the full interest curve \cite{Risk,Hull}. The empirical 
description of the time variation of the interest curve is therefore very 
useful for many practical applications. 

In a previous paper \cite{FRC}, a series of observations concerning the 
U.S. forward rate curve ({\sc frc}) $f(t,\theta)$ were reported, which are in 
strong disagreement with the predictions 
of standard models of interest rates in the literature. 
The most striking empirical results are the following:

\begin{itemize}

\item The average shape of the {\sc frc} is well fitted by a square-root law as
a function of maturity, with a prefactor very close to the spot rate volatility.
This strongly suggests that the forward rate curve is calculated by the money lenders
(who dominate the interest rate markets) using a {\it Value-at-Risk (VaR) like procedure}, and not, 
as assumed in standard models, through an averaging procedure. 
More
precisely, since the forward rate $f(t,\theta)$ is the agreed value at time $t$
of what will be the value of the spot rate at time $t+\theta$, a VaR-pricing 
amounts to writing:
\begin{equation}\label{VaR}
\int^{\infty}_{f(t,\theta)}dr'~P_M(r',t+\theta|r,t)=p,
\end{equation}
where $r$ is the value of the spot rate at time $t$ and $P_M$ is the market 
implied probability of the future spot rate at time $t+\theta$. The value of $p$
is a constant describing the risk-averseness of money lenders. The risk is that 
the spot rate at time $t+\theta$, $r(t+\theta)$, turns out be larger than the 
agreed rate $f(t,\theta)$. This probability is equal to $p$ within the above VaR
pricing procedure. If $r(t)$ performs a simple unbiased random walk, then Eq. (\ref{VaR}) indeed leads to
 $f(t,\theta)=r(t)+A(p)\sigma_r \sqrt{\theta}$, where $\sigma_r$ is the
spot rate volatility and $A(p)$ is some function of $p$.

\item There is a strong {\it positive} correlation between the time variation of the `partial spread' $s(t,\theta) \equiv f(t,\theta)-r(t)$
 and the spot rate $r(t)$, which reaches a maximum around $\theta^*=1$ year and then decreases.
 Within the standard Vasicek model, this correlation is {\it negative} and monotonously decreasing with $\theta$. This
means that a change in $r$ not only results in a parallel translation of the {\sc frc},
but is actually {\it amplified} around $\theta^*$. Correspondingly, the volatility of the
forward rate is found to be `humped' around $\theta^*=1$ year. This can be interpreted within the
above VaR pricing procedure as a time dependent anticipated trend, which is determined
by the past historical trend of the spot rate itself over a certain time 
horizon. In other words, the market looks at the past and extrapolates the observed
trend in the future. When the spot rate goes up, the market anticipated trend also 
typically goes up, and this increase is multiplied by the maturity $\theta$, 
leading to the above mentioned positive correlation.

\item Finally, the decay of the eigenvalues of the deformation correlation matrix for
different maturities, as well as some plausible arguments, suggest that the evolution
equation for $f(t,\theta)$ contains a second derivative, `line tension' term $\partial^2_\theta f(t,\theta)$,
 which tends to smooth out short wavelengths 
deformations of the {\sc frc}. This term is in general absent in arbitrage free models,
but its presence (allowed, for example, by the existence of transaction costs) totally
changes the nature of the equation, and has many interesting consequences \cite{FRC,Cont}. Therefore,
the search for no arbitrage models might be much too strong a constraint to represent
faithfully the empirical behaviour of interest rates. 

\end{itemize}

The aim of the present paper is two fold. First of all, we investigate the empirical
behaviour of the {\sc frc} of five different 
currencies ({\sc usd}, {\sc dem}, {\sc gbp}, {\sc aud} and {\sc jpy}),
 in the period 1987-1999 for the {\sc usd} and 1994-1999 for the other currencies. 
 We discuss the similarities and differences between these currencies. The picture drawn from the {\sc usd} data
is, to a large degree, universal. Second, we specify better the interpretation 
framework sketched in \cite{FRC}, and show how one can calibrate the parameters in a
systematic way. We confirm in particular, the rather spectacular correlation between
the amplitude of the deformation of the {\sc frc} away from its average 
and the past trend of the spot rate.

\section{Data sets and notation}

Our study is based on data sets of daily prices of futures contracts on 
3 month forward interest rates. These contracts and their exchanges are: 
the Eurodollar CME-IMM contract, the Short Sterling LIFFE contract, 
the Euromark LIFFE contract, the Bank Accepted Bills SFE contract and the 
Euroyen TIFFE contract.
In practice, the futures markets price three months forward rates for 
{\it fixed} expiration dates, separated by three month intervals. Identifying three months futures rates to 
instantaneous forward rates, we have available time series on forward rates
$f(t,T_i-t)$, where $T_i$ are fixed dates (March, June, September and December of each
year), which we have converted into fixed maturity (multiple of three months) forward
rates by a simple linear interpolation between the two nearest points such that 
$T_i - t \leq \theta \leq T_{i+1} - t$. 
The shortest available maturity is $\theta_{\min}=3$ months, 
and we identify $f(t,\theta_{\min})$ to the spot rate $r(t)$. 
Table 1 gives the time periods and the number of maturities available for
each data set. We also have analyzed the data corresponding to the {\sc jpy}, for which we only
have 9 different maturities. This last data set is very similar to the {\sc
dem}, which appears to be somewhat different than the other currencies, for
reasons that we shall discuss below. For empirical studies on intraday 
Eurofutures data see Piccinato {\it et-al} \cite{olsen}.

\begin{table}
\begin{center}
\begin{tabular}
{||c|c|c|c||} \hline \hline
& Period & Number of maturities & $\theta_{\max}$ (years) \\ \hline
{\sc usd} 87-99 & 22/6/87-18/2/99 & 11 & 2.75 \\ \hline 
{\sc usd} 94-99 & 1/1/94-18/2/99 & 38 & 9.5\\ \hline
{\sc gbp} & 18/3/94-18/2/99 & 11 & 2.75\\ \hline
{\sc dem} & 15/3/94-22/1/99 & 11 & 2.75\\ \hline
{\sc aud} & 17/3/94-18/2/99 & 11 & 2.75\\ \hline \hline
\end{tabular}
\end{center}
\caption{The datasets}
\end{table}

We will define the `partial' spread $s(t,\theta)$, as the difference between 
the forward rate of maturity $\theta$ and the spot rate: 
$s(t,\theta)=f(t,\theta)-r(t)$. The `long' spread $s(t)$ is 
simply the partial spread at $\theta_{\max}$: $s(t)=s(t,\theta_{\max})$. 
The time evolution of $r(t)$ and $s(t)$, for the different data sets is 
shown in Fig. 1 and Fig. 2. 

Since we only have daily data, our reference time scale will be $\tau=1$
day. The variation of $f(t,\theta)$ between $t$ and $t+\tau$ will be denoted as
$df(t,\theta)$: 
\be
df(t,\theta) = f(t+\tau,\theta)-f(t,\theta).
\ee
More generally, the difference between the values of any observable $O(t)$ between times
$t$ and $t+\tau$ will be noted $dO$. The theoretical time average of $O(t)$ will be 
denoted as $\langle O(t)\rangle$. We will refer to empirical averages (over a finite data set) as 
$\langle O(t)\rangle_e$. For infinite datasets the two averages are the same.

\section{Fundamental Empirical Results}
The purpose of this section is to present the most important 
empirical results that are independent of our chosen {\sc frc} modelisation. 
The details of our model and further important empirical results 
will be discussed in the next 3 sections.

\subsection{The average {\sc frc}}
We consider first the average {\sc frc}, which can be obtained from 
empirical data 
by averaging the partial spread $s(t,\theta)$:
\begin{equation}
\langle s(t,\theta) \rangle_e = \langle f(t,\theta)-r(t) \rangle_e.
\end{equation}

In Figures 3 and 4 we show the average {\sc frc}, $\langle s(t,\theta) \rangle_e$, 
for the
available data sets. As noticed in \cite{FRC} for the {\sc usd} case, these
average curves can be quite satisfactorily fitted by a simple square-root law, 
except for the {\sc dem} (see the bottom left subfigure of Figure 4). For all
these data sets, we also show the following best fit:
\begin{equation}\label{eq35}
\langle s(t,\theta) \rangle_e=a\Bigl(\sqrt{\theta}-\sqrt{\theta_{\min}}\Bigr).
\end{equation}
The corresponding values of $a$ (in $\%$ per $\sqrt{\mbox{year}}$) are given
in Table 2. It is interesting to compare the value of $a/\sqrt{250}$ with 
the daily volatility of the spot rate, which we shall denote $\sigma_r$. 
For all cases, except for the {\sc dem} and the {\sc jpy}, we find that these two 
quantities are very close to each other. 

\begin{table}
\begin{center}
\begin{tabular}
{||c|c|c|c||} \hline \hline
& $a$ & $\sigma_r$ & $a/\sqrt{250}$ \\ \hline
{\sc usd} 94-99 & 0.78 & 0.047 & 0.049 \\ \hline 
{\sc usd} 87-99 & 1.18 & 0.067 & 0.075\\ \hline
{\sc gbp} & 0.91 & 0.053 & 0.058\\ \hline
{\sc aud} & 1.43 & 0.078  & 0.090 \\ \hline
{\sc dem} & 1.50 & 0.033 & 0.095 \\ \hline \hline
\end{tabular}
\end{center}
\caption{Parameters for Eq. (\protect\ref{eq35}). The units for $\sigma_r$ are
\% per square-root day.}
\end{table}

We thus fully confirm here -- with much more empirical data -- the suggestion 
of ref. \cite{FRC} that the
{\sc frc} is on average fixed by a VaR-like procedure, specified by Eq. (\ref{VaR}) above.
The {\sc frc} appears to be the {\it envelope} of the
future evolution of the spot rate. If the spot rate fluctuates around a reasonable 
equilibrium value, then the anticipated trend on the spot rate is {\it on average} zero.
The implied probability $P_M$, appearing in Eq.  (\ref{VaR}) is therefore centered around
the present value of the spot rate; this leads to the simple square-root law discussed above.

As will be discussed at length in the next sections, even when the anticipated trend is on average
zero, it is not zero for any given day. Its presence leads to much of the interesting effects seen on the deformation of the {\sc frc} 
away from its average shape. There are however, as we discuss now, situations where the average
anticipated trend is expected to be non zero.

\subsection{Mean reversion effects} \label{dem}

As obvious from Figure 4 and Table 2, the {\sc dem} data is different. The {\sc frc} 
increases with maturity much faster than it should according to the unbiased VaR procedure
explained above. However, a look at Figures 1 and 2 allows one to understand why this might be so. 
In all the cases, except the {\sc dem}, the spot rate is fluctuating around an `equilibrium' rate of about 6.5\%. 
Interest rates are neither too low nor too high 
and the markets have no reason to expect the future value of the spot rate to move outside a fairly 
narrow band. On the other hand the {\sc dem} spot rate shows a large 
down trend and the average spot is a low $3.93\%$ 
(a similar but more extreme situation is seen on the {\sc jpy} data). 
It is not in an equilibrium regime, or at least the markets do 
anticipate a reversion towards higher rates.\footnote{The expectation of a rise of the spot
rate might also be due in part to the `Euro effect'.} One might then expect that for 
`short' time periods, where the spot rate remained low or high on average, 
the shape function 
$Y(\theta)$ can  effectively be spot dependent. A reasonable guess, inspired
from the Vasicek model, is that 
the partial spread $s(t,\theta)$, evolves about a time dependent mean, given by:
\begin{equation}\label{eq33}
s_0(r,\theta)= \sigma_r\left(\sqrt{\theta}-\sqrt{\theta_{\rm min}}\right)
+\Bigl(r_0-r(t)\Bigr)\left(1-e^{-\lambda_r(\theta-
\theta_{\rm min})}\right),
\end{equation}
which is the sum of the VaR-like term $\sqrt{\theta}$ and a mean reverting 
term, characterized by an equilibrium spot rate $r_0$ and a reversion time
$1/\lambda_r$. 
A consequence of Eq. (\ref{eq33}) is that the empirical average of the 
partial spread becomes:
\begin{equation}\label{eq34}
\langle s(t,\theta)\rangle_e= \sigma_r\left(\sqrt{\theta}-\sqrt{\theta_{\rm min}}\right)
+\Bigl(r_0-\langle r(t)\rangle_e\Bigr)\left(1-e^{-\lambda_r(\theta-
\theta_{\rm min})}\right).
\end{equation}
For $\langle r(t)\rangle_e \simeq r_0$, the mean reversion term is absent, 
and one recovers the simple $\sqrt{\theta}$ law of Eq. (\ref{eq35}). 
In the bottom right of Figure 4 we have found the best fit of 
$\langle s(t,\theta)\rangle_e$ 
based on Eq. (\ref{eq34}). Using the empirical {\sc dem} volatility for $\sigma_r$, 
we found $\lambda_r=0.22$ per year and $r_0-\langle r \rangle_e=2.98 \%$.
Since for the {\sc dem} data $\langle r(t)\rangle_e=3.9\%$, these results 
imply that the market expects a mean reversion to approximately 
$7\%$ over a timescale of 5 years. These numbers are very realistic and 
suggest that Eq. (\ref{eq34}) qualitatively accounts for the strong anticipated upward
trend present in the {\sc dem} data.  Of course, 
for datasets long compared to the spot mean reversion timescale, 
the empirical average reduces to the model average and 
Eqs. (\ref{eq35}) and (\ref{eq34}) are the same.
Much the same features are found for the {\sc jpy}.

\subsection{Spread-spot response function and volatility of the {\sc frc}}

We shall study  how a change in the value of the spot rate affects the rest of the {\sc frc},
after subtracting the trivial overall translation of the curve. This will define a 
`spread-spot response function'\footnote{here after we call this simply `the response
function'}  ${\cal R}(\theta)$, through the following dynamical equation:
\begin{equation}\label{eqresponse}
df(t,\theta)=[1+{\cal R}(\theta)]dr(t)+dz(t,\theta),
\end{equation}
where, $dz(t,\theta)$, are spot independent noise increments and 
we have the constraints:
\begin{equation}
{\cal R}(\theta_{\rm min})=dz(t,\theta_{\rm min})=0.
\end{equation}
If the response function is positive, then a change in the spot is {\it amplified}
along the {\sc frc}. As we discuss below, the response function is, in fact, a more fundamental quantity than the {\sc frc} 
volatility. It determines much of the qualitative shape of the {\sc frc} volatility
which, itself, is given by a less intuitive contribution of the three terms 
appearing in Eq. (\ref{eqresponse}).

From Eq. (\ref{eqresponse}) we find that the response function, ${\cal R}(\theta)$, 
can be measured by computing the empirical average:
\begin{equation}
{\cal R}(\theta)=\frac{\langle ds(t,\theta)dr\rangle_e}{\langle dr^2\rangle_e}.
\end{equation}
In figure 5 we show, for two time periods, the response function
for our {\sc usd} data, while in figure 6 we show the response function
for three different currencies, all for the same time period.
We see that in all cases, ${\cal R}(\theta)$ reaches a positive peak at 
$\theta=0.75$ -- $1$ year, followed by a decay to some negative value. 
The same features were found for the {\sc jpy}.
This qualitative shape of the response function appears to be rather {\it universal} 
across currencies and periods. In figure 9 we show the empirical 
volatility for the {\sc usd}, defined as:
\begin{equation}
\sigma(\theta)=\sqrt{\langle df^2(t,\theta)\rangle_e}.
\end{equation}
We see a strong peak in the volatility at 
1 year \cite{Hull,Moraleda}. For the {\sc usd} 90-93 inclusive, which is not shown, we find a similar 
peak though not as strong. In figure 10 we show the empirical volatility  for the 
other currencies. In all cases the volatility shows a steep initial {\it rise} 
as the maturity grows. We will demonstrate in the next section that this
robust qualitative feature can be traced to the positive peak found 
for the response function.

While it seems reasonable to to expect the response function to become small 
for longer maturities, it is not {\it a priori} clear why it 
should show a robust positive peak around 1 year. This is actually in stark
contrast to the Vasicek model \cite{Risk,Hull}, where ${\cal R}$ is found to be 
{\it negative}, and monotonously decreasing. Correspondingly, the volatility is 
exponentially decaying 
with maturity. Previously \cite{FRC} it was 
proposed that the shape of the response function, and consequently that of the 
{\sc frc} volatility, could be explained by an anticipated trend, calibrated on the past
behaviour of the spot. Qualitatively, 
when the market tries to estimate the implied probability $P_M(r',t+\theta|r,t)$, it looks at the past spot trend over 
some time horizon and extrapolates this trend into the future. In sections 5 and 6 
we will present striking empirical confirmation of this proposal.

\section{A Phenomenological Model}
Based on our empirical findings \cite{FRC}, we consider the following 
decomposition of the {\sc frc} dynamics:
\begin{equation}
f(t,\theta)=r(t)+s(t)Y(\theta)+y(t,\theta),
\end{equation} 
with the following constraints:
\begin{equation}
Y(\theta_{\rm min})=y(t,\theta_{\rm min})=0,
\;\;Y(\theta_{\rm max})=1,\;\;
y(t,\theta_{\rm max})=0.
\end{equation}
We also impose that:
\begin{equation}\label{eq43}
\langle y(t,\theta)\rangle=0.
\end{equation}
We find from Eq. (4.1) that the average {\sc frc} is given by:
\begin{equation}
\langle s(t,\theta) \rangle= \langle f(t,\theta)-r(t) \rangle =
\langle s(t) \rangle Y(\theta).
\end{equation}
The function $Y(\theta)$ can therefore, up to a scaling by the average long spread, 
be interpreted as the average {\sc frc}, and can be calibrated to the empirical 
average {\sc frc} discussed in section 3.1. 
Clearly the `deformation process' $y(t,\theta)$ now describes the fluctuations 
around the average {\sc frc}.

We now turn to a more precise description of the {\sc frc} decomposition 
Eq. (4.1).
In order to rationalise the data we propose to think of the evolution of 
the {\sc frc} as driven by three independent noises $du, dv$ and $dw$, of unit variance:
\begin{equation}
\langle du^2\rangle=1,\;\;\langle dv^2\rangle=1,\;\;\langle dw^2\rangle=1,
\end{equation}
and 
\begin{equation}
\langle du~dv\rangle=0,\;\;\langle du~dw\rangle=0,\;\;\langle dv~dw\rangle=0.
\end{equation}
These three random factors affect the spot rate, the `long' spread and the deformation as follows. The spot rate is the most important factor which, as will be shown below,
also strongly affects the evolution of the deformation $y(t,\theta)$. We define:
\be
dr(t)=\sigma_r du, 
\ee
where $\sigma_r$ is the spot rate volatility. Although our results are compatible with 
more general models for the  `long spread', we choose for definiteness the following mean reverting equation:
\begin{equation}
ds(t)=\lambda_s\Bigl(s_0(r)-s(t)\Bigr)dt+\sigma_r(\mu du+\nu dv),
\end{equation}
where $s_0(r)$ is the (possibly spot dependent -- see  subsection \ref{dem}) 
reversion level and $\lambda_s$ describes the 
mean reversion speed.  For the measurements described in this paper, 
however, the precise form of the spot and long spread mean reversion terms are
not important since they are in general negligible compared to the 
noise terms. The coefficient $\mu$ measures the influence 
of the spot rate on the long spread. 
More precisely:
\begin{equation}\label{dsdr}
\langle dsdr\rangle=\sigma^2_r\mu.
\end{equation}
The spread volatility is given by:
\begin{equation}\label{ds2}
\langle ds^2\rangle=\sigma^2_r\Bigl[\mu^2+\nu^2\Bigr].
\end{equation}
Using Eqs. (\ref{dsdr}) and (\ref{ds2}), one can determine the 
coefficients $\mu$ and $\nu$ from the data. These coefficients are 
shown in Table 3.
\begin{table}[b]
\begin{center}
\begin{tabular}
{||c|c|c|c||} \hline \hline
& $\mu$ & $\nu$ \\ \hline
{\sc usd} 94-99 & -0.22 & 1.17 \\ \hline 
{\sc usd} 87-99 & -0.18 & 0.65 \\ \hline
{\sc gbp} & -0.18 & 1.11 \\ \hline
{\sc aud} & -0.05 & 0.99  \\ \hline
{\sc dem} & 0.02 & 1.22 \\ \hline \hline
\end{tabular}
\end{center}
\caption{Parameters $\mu$ and $\nu$ extracted from Eqs. (4.9) and (4.10).}
\end{table}
We see from this table and Eq. (4.8) that the long spread 
evolves mostly independent from the spot, but there is a persistent 
negative correlation between the spot and long spread. 
Note also that except for 
{\sc usd} 94-99, the spread uses the 2.75 maturity forward rate. For 
longer maturity spreads we can expect stronger negative correlations.
(Figures 5 and 6 for the response function suggest that the 
negative response grows slowly with maturity.)

As for the `deformation process' $y(t,\theta)$, our analysis of the empirical 
data leads us to write it as:
\begin{equation}\label{yy}
y(t,\theta)=C(\theta)b(t)+G(\theta)\eta(t), 
\end{equation}
with the constraints:
\begin{equation}
C(\theta_{\rm min})=G(\theta_{\rm min})=0,
\;\;C(\theta_{\rm max})=G(\theta_{\rm max})=0.
\end{equation}
The function $b(t)$ is what was called in \cite{FRC} the `anticipated trend', 
while $\eta(t)$
is a mean reverting noise term. Due to Eq. (4.3), we require that:
\begin{equation}
\langle b(t)\rangle=\langle \eta(t)\rangle=0.
\end{equation}
The logic behind Eq. (\ref{yy}) is the following: within a VaR-like pricing, the {\sc frc}
is the envelope of the future anticipated evolution of the spot rate. On average, this
evolution is unbiased (provided $r(t)$ is not too far from its equilibrium value $r_0$),
and the average {\sc frc} is a simple square-root. However, at any particular time $t$, the market 
anticipates a future trend $b(t)$. This means that the probability distribution of the spot,
$P_M(r',t+\theta|r,t)$ is not centred around $r$, but around $r + b(t)\theta$ (for small
$\theta$). However, the market also `knows' that its estimate of the trend will not
be good on the long run. We therefore write more generally that the distribution is
centered around $r + b(t)C(\theta)$. 

How does the market determine $b(t)$? One of the main proposals of \cite{FRC} was that the
anticipated trend actually reflects the actual past trend of the spot rate. In other words, 
the market extrapolates the observed past behaviour of the spot to the nearby future. More 
precisely, we write:
\be\label{kernel}
b(t)= \int_{-\infty}^t  K(t-t') d r(t'),
\ee
where $K$ is an averaging kernel, which we normalize such that $K(0)=1$. We have considered two choices
for $K(u)$: an exponential form, which leads to an Ornstein-Uhlenbeck ({\sc ou}) process for $b$ driven
by the spot rate $r(t)$,
\begin{equation}\label{db}
db(t)=-\lambda_{b} b(t) dt+ dr(t),
\end{equation}
or a simple flat window ({\sc fw}) of width $T_b$, leading to:
\be
b(t)={r(t)-r(t-T_b)}.
\ee
At this stage, the model has only two factors: the spot rate, and the noise term $dv$ driving the long spread.
 It is clear that since the long term extrapolation of the anticipated trend is doubtful, one should also add an
 extra noise term $G(\theta) \eta(t)$ to $r + b(t)C(\theta)$, where the
`signal to noise' ratio $C(\theta)/G(\theta)$ should become small for large $\theta$'s.
For the noise term $\eta(t)$ we also propose a simple mean reverting ({\sc ou}) process driven by our third factor $dw$:
\begin{equation}\label{eta}
d\eta(t)=-\lambda_{\eta} \eta(t) dt+\sigma_r dw.
\end{equation}
We have thus assumed 
that the deformation is driven by the spot and a noise. Because the noise $dw$ has been 
scaled by $\sigma_r$, we can directly compare the functions $G$ and $C$. 
By comparing them we can see how much of the deformation is driven by the spot compared 
to noise.  The model assumption in (\ref{eta}) is that the deformation noise is 
furthermore independent of the long spread noise. 
Eq. (\ref{eta}), together with the
preceding equations, fully defines our three factor model. 

In our model the response function is given
by:
\begin{equation}
{\cal R}(\theta)=\frac{\langle ds(t,\theta)dr\rangle}{\langle dr^2\rangle} =\mu Y(\theta)
+C(\theta).
\end{equation}
Note that for $\theta_{\max}$, this equation is identical to Eq. (\ref{dsdr}) above. 
From the previous knowledge of $Y(\theta)$ and $\mu$,
we can calibrate $C(\theta)$ to our empirical response function, 
${\cal R}(\theta)$, discussed in section 3.3. The results are shown in 
Figures 7 and 8. We again find a robust positive peak at $\theta=0.75$ -- $1$ year 
which is due to the same positive peak found in figures 5 and 6 
for the response function.

Neglecting the contribution of all drifts, the square of the maturity dependent 
volatility of the forward rates read, within our theoretical framework:
\begin{eqnarray}
\sigma^2(\theta) &=&
\sigma^2_r\left[1+2C(\theta)+2\mu Y(\theta)
+Y^2(\theta)(\mu^2+\nu^2) \right. \nonumber \\
&+& \left. 2\mu Y(\theta)C(\theta)
+Q_K C^2(\theta)+G^2(\theta)\right],
\end{eqnarray}
where $Q_K=1$ when the averaging kernel defining $b(t)$ is taken to be exponential, 
and $Q_K=2$ if it is taken to be flat. 
For $\theta_{\max}$, this equation reduces to 
the previous Eq's (\ref{dsdr},\ref{ds2}). 
The only unknown left to calibrate using this equation is 
the noise contribution to the deformation, $G(\theta)$. 

In Figs. 9 and 10 we plot the empirical volatility and the square root of Eq. (4.19) 
when $C=G=0$ and also when $G=0$. In this section we have chosen to use the value $Q_K=2$ since this 
leads to a higher signal to noise ratio. Another justification of this choice will be given 
in the next section. 
For the former case, the deformation has been set to zero, and the model 
has only been calibrated to the spot, the long spread volatility 
and the average {\sc frc}. In the latter case the model is also calibrated 
to the response function which includes the effect of the proposed anticipated 
bias (but not the 3rd noise factor). 
The empirical {\sc frc} volatility was discussed in section 3.3 where we emphasized the 
universal steep initial rise in the volatility with increasing maturity.
We can now see that the response function, and consequently the proposed 
anticipated bias, make a large contribution to the {\sc frc} volatility and, in 
fact, seem to be be responsible for the universal steep rise in the {\sc frc} volatility 
function. Using Eq. (4.19), we can calibrate $G(\theta)$ to the empirical volatility 
shown in Figs. 9 and 10. 
The signal to noise ratio $C(\theta)/G(\theta)$ is plotted in Figure 11 for the 
different currencies. It is seen to decrease monotonously with $\theta$, in accordance 
with our interpretation of $C(\theta)$ as an extrapolated trend contribution, the 
influence of which is bound to decay with maturities.

We have chosen to describe the {\sc frc} using three (correlated) components that have
a direct financial interpretation: the spot, long spread and deformation. It is 
useful to compare the volatility and kurtosis of these three components. For the deformation, 
we consider the 1 year maturity, corresponding to the maximum amplitude of this component. 
We show the results in Table 4 below. It is interesting to see the high kurtosis of all
components and the relative stability of the deformation volatility across the 
different datasets. Even though the long spread volatility is greater than the deformation 
volatility, the deformation remains the more important component at the short end of 
the {\sc frc}. This follows because of the function $Y(\theta)$ goes to zero 
for short maturities (see Eq. (4.1)).

\begin{table}
\begin{center}
\begin{tabular}
{||c|c|c|c|c|c|c||} \hline \hline
& $\sigma_r$ & $\sigma_s$ & $\sigma_y$ & $k_r$ & $k_s$ & $k_y$ \\ \hline
{\sc usd} 94-99 & 0.047 & 0.051 & 0.035 & 6.6 & 6.9 & 8.4 \\ \hline 
{\sc usd} 87-99 & 0.067 & 0.045 & 0.028 & 37.8 & 9.7 & 6.2 \\ \hline
{\sc gbp} & 0.053 & 0.059 & 0.029 & 5.9 & 10.6 &  12.9\\ \hline
{\sc aud} & 0.078 & 0.078  & 0.028 & 8.0 & 6.4 & 5.6 \\ \hline
{\sc dem} & 0.033 & 0.040  & 0.028 & 5.5 & 3.2 & 4.3\\ \hline \hline
\end{tabular}
\end{center}
\caption{Daily volatility and kurtosis of the spot, long spread and deformation (1 year 
maturity) for the datasets. The volatility units are \% per square-root day.}
\end{table}

\section{Direct Confirmation of the Anticipated Trend}

Both in  \cite{FRC} and in the previous section, the anticipated trend, defined by Eq. (4.14),
was proposed as an explanation of the robust positive peak in the 
response function, Eq. (4.18). We also saw that the 
anticipated trend could explain the qualitative shape of all the {\sc frc} volatility 
functions as well as the decay in the signal to noise ratio. 
However, although this empirical evidence is 
compelling, it does not constitute the most direct empirical confirmation of the 
anticipated trend. This is because all the previous 
measurements are increment based and not directly  sensitive to the 
memory timescale of the averaging kernel in 
Eq. (4.14).

To more directly confirm the anticipated trend mechanism,  
we need to study the empirical deformation process itself rather than 
the {\sc frc} increments. The empirical deformation process is easily extracted from 
the data using:
\begin{equation}
y(t,\theta)=f(t,\theta)-r(t)-s(t)Y(\theta).
\end{equation} 
By construction it is zero at the endpoints and also has 
zero mean. In our current 3 factor formulation, our model 
deformation is described by Eq. (4.11) where we consider 
Eq's. (4.15) and (4.16) as possible models for the anticipated trend $b(t)$,
and Eq. (4.17) as a simple model for the noise component. 
We saw in the previous section how the functions $C(\theta)$ and $G(\theta)$ 
can be calibrated using the response function, Eq. (4.18), and 
the volatility, Eq. (4.19). To determine the model parameters $\lambda_{\eta}$ and 
$T_b$ or $\lambda_{b}$, we propose to measure the following average error: 
\begin{equation}
\sqrt{\Bigl< \Bigl(y(t,\theta)-C(\theta)b(t)\Bigr)^2\Bigr>}=G(\theta)\sqrt{\langle 
\eta^2(t)\rangle}=\frac{G(\theta)\sigma_r}{\sqrt{2\lambda_{\eta}}},
\end{equation}
which follows from Eq's. (4.11) and (4.17).
To measure the {\sc lhs} of the above equation (the average error), 
we must first extract the empirical deformation $y(t,\theta)$ using Eq. (5.1). 
We then determine $b(t)$ using the empirical spot time series and 
Eq. (4.15) or (4.16).\footnote{In this empirical determination of $b(t)$ 
we actually use detrended spot increments, defined as 
$d\hat r(t)=dr(t)-\langle dr \rangle_e $.} 
The {\sc lhs} of Eq. (5.2) will have a minimum error for some $T_b$ or $\lambda_b$. 
This is the timescale where the deformation and anticipated trend 
match up best, thereby fixing the values of $T_b$ or $\lambda_b$. We can then 
use the magnitude of this minimum error to fix $\lambda_{\eta}$ via 
Eq. (5.2). 

Consider first the flat window model, Eq. (4.16) ({\sc fw} model). 
In Figure 12 we plot the {\sc lhs} of Eq. (5.2) for {\sc usd} 94-99, against the 
parameter $T_b$, used in the simulation of $b(t)$. We consider the first 4 maturities 
beyond the spot-rate. 
As $T_b$ becomes small, the anticipated trend 
becomes negligible compared to the deformation in the {\sc lhs} of Eq. (5.2). In this
limit the \% error in Figure 12 reduces to the deformation root mean square deviation. 
Let us first assume that the market believes that the past spot trend contains 
no useful information about the future spot-rate - in other words, the market 
believes in the efficient market hypothesis.
In this case the deformation process should be uncorrelated with the anticipated trend. 
This in-turn implies that the {\sc lhs} of Eq. (5.2) should increase with $T_b$.
In fact, what one sees is a deep minimum indicating the clear presence of a 
dynamical timescale around $100$ days. This demonstrates that the deformation and anticipated trend 
are strongly correlated.
In Figure 13 we plot 
the empirical deformation against $C(\theta)b(t)$, where we have 
set $T_b=100$ days. Indeed, we visually confirm a striking correlation, even out to the 
15 month maturity. In Figure 14 we plot the {\sc lhs} of Eq. (5.2) 
for the {\sc gbp}, {\sc dem} 
and {\sc aud} case as a function of $T_b$. We again find minima at $175$, $100$ and $130$ 
days respectively. In these cases the minima are not so deep which makes the timescale 
not as well defined as in the {\sc usd} case. 
In Figure 15 we plot for the {\sc gbp}, {\sc dem} and {\sc aud}, the 
deformation, $y(t,\theta)$, against $C(\theta)b(t)$, for $\theta$= 6 months. 
We again see a clear correlation 
though the quality is not as good as for the {\sc usd} 94-99 case. This 
was expected since the signal to noise ratio $C/G$ is smaller in these cases. 
However the same qualitative features are present for all datasets.

Using Eq. (5.2) we can now estimate $\lambda_{\eta}$. We use the value of the
maturity $\theta$ such that $G(\theta)$ is maximum: $\theta=3$ years for the
{\sc usd} 94-99 data and $\theta=1.25$ years for other currencies. The 
values of the different time scales (in trading days) are reproduced in Table 5.
\begin{table}
\begin{center}
\begin{tabular}
{||c|c|c||} \hline \hline
& $T_b$ & $\lambda_{\eta}^{-1}$ \\ \hline
{\sc usd} 94-99 & 100 & 151 \\ \hline 
{\sc gbp} & 175 & 169  \\ \hline
{\sc aud} & 130 & 114  \\ \hline
{\sc dem} & 100 & 104   \\ \hline \hline
\end{tabular}
\end{center}
\caption{Table of the memory time $T_b$ for the {\sc fw} trend model (4.16)
and the noise mean reversion time extracted from Eq. (5.2). The units are 
trading days.}
\end{table}
As a rule of thumb we can say that the deformation dynamics is mean
reverting on a time scale of $\simeq 6$ months. It must be emphasized 
that these results do not constitute a statistical validation of the drift term 
in, Eq. (4.17), for the deformation noise. The mean reverting nature of 
this model was proposed in relation to the existence of a second order derivative term
in the evolution of the {\sc frc}, discussed in  \cite{FRC,Cont}. The specific results obtained for $\lambda_{\eta}$ are 
only meaningful if the mean reversion assumption is correct.

We have performed an identical analysis using the {\sc ou} model for the 
anticipated trend. We find the same optimal timescales as for the {\sc fw} model. 
We find that the minimal errors for the {\sc gbp} and {\sc aud} are larger for this 
model, the error for the {\sc dem} is similar but the error for the {\sc usd} is 
less (by about 20\%). Even though the {\sc usd} error is less within a {\sc ou} 
model, overall we prefer the {\sc fw} model in a three factor setting for the following 
reasons:

\begin{itemize}

\item It leads to deeper minima and therefore to a better defined optimal timescale

\item More significantly, the contribution to the {\sc frc} volatility of the anticipated
trend is higher, through the factor $Q_K$ defined in Eq. (4.19). 
This in turn reduces the value of $G(\theta)$ which then increases the signal to 
noise ratio. For the {\sc usd} case the reduction of $G$ is rather high for the 
early maturities. This point is significant because it means we are much better 
able to simultaneously fit the {\sc frc} volatility and response function by 
$C(\theta)$ alone.

\item From Eq. (5.2), a smaller $G(\theta)$ then leads to a larger mean 
reversion timescale for the noise. For the {\sc usd} data the {\sc ou} model 
gives $\lambda_{\eta}^{-1}\simeq 50$ days for the first 4 maturities -- somewhat lower than that obtained for the {\sc fw} model.

\item Once the optimal $T_b$ has been obtained using the 
above method, it is interesting to calibrate $C(\theta)$ such that 
the {\sc lhs} of Eq. (5.2) is minimal. This of course means that we do not try {\it a priori} to 
fit the response function exactly. We find that this method does not lead to any 
significant reduction of the error. This is a further nice feature of the {\sc fw} model: the 
function $C(\theta)$ obtained from the response function 
also seems to give the optimal fit to the deformation.

\end{itemize}

\section{Reducing the Number of Factors}

For the {\sc usd} 94-99 case, the high signal to noise ratio shown in Fig. 11 
suggests that the short end of the {\sc frc} could be rather well modelled by a 
simpler one or two-factor version of our model. 
One way to obtain the latter is by simply setting $G(\theta)=0$. 
In these cases we can not calibrate our model to both the response function,
Eq. (4.18), and the volatility, Eq. (4.19). 
Since the volatility is generally the more important quantity for applications, 
we choose here to calibrate $C(\theta)$ directly to Eq. (4.19) with $G(\theta)=0$. 
In the previous section we calibrated $C(\theta)$ to the response function. In this 
case the result obtained is independent of our choice of model for the anticipated 
trend. 
This is no longer the case when we calibrate $C(\theta)$ to the volatility. 
Using this calibration procedure for the {\sc usd} 94-99, it turns out that 
the match between the deformation and anticipated trend, for the {\sc fw} model, 
is similar to that obtained in section 5. On the other hand we find the match is
improved when we use the {\sc ou} model for the anticipated trend. 

In Table 6 below, we compare $C(\theta)$ fitted to the 
response function (as in the previous sections) and $C(\theta)$ fitted 
using the volatility in the two-factor version of our model, where we have 
now chosen the {\sc ou} model for the anticipated trend. 
We also compare the minimum errors (at 100 days) for the previous {\sc fw} model 
and now for the {\sc ou} model using the new values for $C(\theta)$. 
For the {\sc ou} model a plot similar to Figure 12 shows again a 
minimum at $\lambda_b^{-1}=100$ days.

\begin{table}[b]
\begin{center}
\begin{tabular}
{||c|c|c|c|c||} \hline \hline
& $C(\theta)$ (3 fac.) & $C(\theta)$ (2 fac.)& Err --  {\sc fw} 3 fac. & Err -- {\sc ou} 2 fac. \\ \hline
6 months & 0.32 & 0.36 & 0.10 &0.074\\ \hline 
9 months & 0.48 & 0.55 & 0.16 & 0.11\\ \hline
12 months & 0.50 & 0.61 & 0.17 & 0.12\\ \hline
15 months & 0.48 & 0.60 & 0.16 & 0.12\\ \hline \hline
\end{tabular}
\end{center}
\caption{Comparing the fit to the deformation for the 
3 factor and 2 factor models. For the {\sc usd} 94-99.}
\end{table}
In Figure 16 we plot both the deformation and anticipated trend 
using the {\sc ou} model. The correlation is quite spectacular and 
completely validates the anticipated trend proposal. Clearly, when 
the signal to noise ratio is high the two-factor version is an 
excellent model of the {\sc frc} deformation.
This two-factor model is not as effective for the other currencies 
since the signal to noise ratio is worse.

For the {\sc usd} 94-99 data set, not only is the signal to noise 
ratio much higher than for the other currencies, but the signal 
to noise ratio persists for longer maturities. From Figure 11 we 
see that the $2.25$ years maturity forward rate has a signal to noise 
ratio of $\simeq 0.5$. For the other currencies this occurs at approximately 
the $0.75$ year maturity. In Figure 17 we plot the deformation and 
trend for the $2.25$ year ($27$ months) maturity for both our three-factor and 
two-factor models. The strong correlation is clearly seen to persist 
even beyond 2 years forward of the spot! 

For the {\sc usd} 94-99 case, it is important to note that even 
the one factor version of our model provides a very good description 
of the deformation about the average curve at the short end. 
The one factor version of our model is simply:
\begin{equation}
f(t,\theta)=r(t) +a\Bigl(\sqrt{\theta}-\sqrt{\theta_{\min}}\Bigr)
+{\cal R}(\theta)b(t)
\end{equation}
where $a\simeq \sigma_r$ and ${\cal R}(\theta)$ is the response function. 
However, as before, we may prefer to 
calibrate ${\cal R}(\theta)$ to the volatility which now takes the form:
\begin{equation}
\sigma^2(\theta) =
\sigma^2_r\left[1+2{\cal R}(\theta)+Q_K {\cal R}^2(\theta)\right].
\end{equation}
This model works very well at the short end because, for the early maturities, 
the spot trend contribution is much greater than the long spread 
contribution in setting the {\sc frc}. 
This point is particularly important since the short end of the curve 
is the most liquid part of the curve, corresponding to the largest volume of
trading (in particular on derivative markets). Interest rate models have evolved
towards including more and more factors to get better 
models. Yet our study suggests that after the spot, it is the {\it spot 
trend} which is the most important model component. 

It is important to understand if the popular {\sc hjm} framework \cite{hjm}
can capture the qualitative features of the simple model Eq. (6.1). 
Significantly, it seems clear to us that our empirical results for 
the average {\sc frc} cannot be naturally accounted for in this framework. 
For the one-factor {\sc hjm} model, the relation between the volatility and response 
function is the same as Eq. (6.2), when $Q_K=1$. Empirically we find that 
a better simultaneous fit of the volatility and response function is obtained 
when $Q_K=2$. What is more important but less clear, 
is whether the {\sc hjm} framework captures the 
striking spot trend effect observed. The problem here is that the 
complexity of the {\sc hjm} framework does not allow us to easily 
see how the spot rate drives the forward rates. 
On the other hand, our model 
is very close in spirit to the strong correlation limit of the `two-factor' 
spot rate model of Hull-White \cite{hull2}, 
which was introduced in an {\it ad hoc} 
way to reproduce the volatility hump. 
Although phrased differently, this model assumes in effect the existence of an
extrapolated trend following an {\sc ou} process driven by the 
spot rate. In the one-factor perfect correlation limit of this model, we can show 
that up to negligible $\sigma^2$ terms, the model leads to:
\be
f(t,\theta)=r(t)+b(t) {\cal R}_{HW}(\theta),
\ee
where $b(t)$ is given by Eq. (\ref{db}), while ${\cal R}_{HW}(\theta)$ is obtained explicitly as:
\be\label{HW2}
{\cal R}_{HW}(\theta)= \frac{\lambda_r}{\lambda_r-\lambda_b} \left(e^{-\lambda_b \theta}
-e^{-\lambda_r\theta}\right).
\ee
One can check that ${\cal R}_{HW}(\theta)$ is positive and has a correct qualitative
shape, growing linearly for small maturities and decaying to zero for
large maturities. This `Hull-White II' model however fails to capture the $\sqrt{\theta}$ 
behaviour of the average {\sc frc}. We furthermore believe that our analysis 
gives a rather compelling financial interpretation of their model.

\section{Discussion and conclusion}

In this paper, we analyzed in detail the empirical properties of the forward rate curve corresponding to different currencies. 
We proposed a theoretical framework to account for the observed properties which builds upon the intuitive picture
 proposed in \cite{FRC}, and showed in detail how the different parameters entering the model can be calibrated using
historical data. The main conclusions of our work are the following:
\begin{itemize}
\item The average {\sc frc} indeed follows a simple square-root law, with a prefactor 
closely related to the spot volatility. This strengthens the idea of a 
VaR-like pricing of the
{\sc frc} proposed in \cite{FRC}. When the spot rate is very low (or very high) compared to its historical average, 
this square-root law fails to describe the data. One can indeed 
expect a strong average anticipated mean reversion contribution in these cases. This allows to rationalize our 
findings on the {\sc dem} (and {\sc jpy}) data.

\item The {\sc frc} is strongly correlated with the past trend 
observed on the spot rate. This is strikingly illustrated in Figures 13, 16 and 17 for 
the {\sc usd} data, where one sees that the
past trend on the spot tracks very closely the amplitude of the 
{\sc frc} deformation. This
correlation allows us to account for the strong positive response function 
and for the volatility hump.

\item 
After the spot, it seems that the {\it spot trend} is the most 
important component for modeling the {\sc frc}.

\end{itemize}

It is important to understand whether the popular {\sc hjm} framework 
can describe the above observed empirical properties. 
We feel that our empirical results for the average {\sc frc} cannot be 
naturally accounted for in this framework. 
For the spot trend effect the situation is less clear. From the empirical 
evidence presented here, it is clear that this term should be included in 
any model of the {\sc frc}. This offers the interesting possibility 
of strongly constraining the set of reasonable {\sc hjm} models. 
We know of only one spot-rate model that does, in effect, include a 
spot trend term. This is the strong correlation limit of the 
Hull-White two-factor model \cite{hull2}. 
However, the precise relation, if any, between this model and the 
{\sc hjm} framework is unclear and needs to be understood. 
For work in this direction see Chiarella and Kwon \cite{kwon}.

It would be interesting to understand from first principles the shape of the 
basic object responsible for the {\sc frc} deformation, namely the 
function $C(\theta)$ which, at the short end, is mainly determined 
by the response function. A reasonable fit is \cite{FRC} 
$C(\theta)=c\theta\exp(-\gamma\theta)$ (see also (\ref{HW2}) above).
The initial linear rise is related to the interpretation of $C(\theta)b(t)$ as an
extrapolated trend; the following exponential decay reflects how the market perceives the
persistence of this trend over time. 
Interestingly, a very similar shape for the structure of the deformation was proposed recently by R. Cont \cite{Cont}, based on the diagonalisation
of an evolution operator for the {\sc frc} which includes a second-order, line tension term,
first proposed in \cite{FRC}. The precise relation -- if any -- between this two view points needs to be clarified.

A natural extension of our work is to adapt the general 
method for option pricing in a non-Gaussian world 
detailed in \cite{JPMP} to interest rate derivatives. 
Work in this direction is in progress.

\vskip 1cm
Acknowledgements:
\vskip 0.5cm

We thank J. P. Aguilar, P. Cizeau, R. Cont, O. Kwon, L. Laloux, M. Meyer, A. Tordjman and in particular M. Potters for many interesting discussions.

\begin{figure}[tbp]
\epsfxsize=14cm
\centering{\ \epsfbox{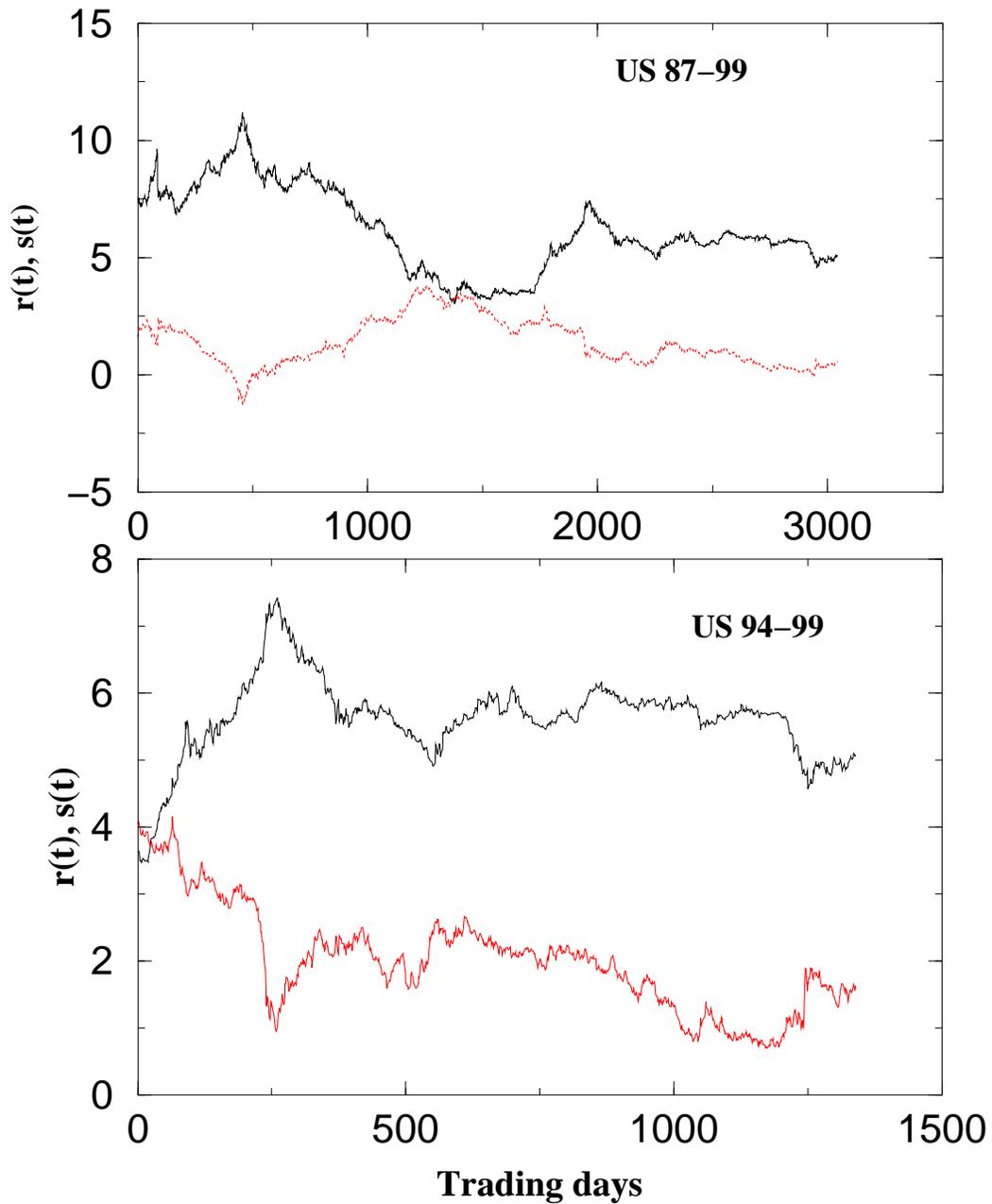}}
\vspace{0.0cm}
\caption{Top Figure: The historical times series of the spot rate (top bold curve)
and long spread for our {\sc usd} 87-99 dataset. Bottom Figure: The same as before 
but now the dataset is the {\sc usd} 94-99. For this later case the long spread is defined 
using the maximum forward rate of 9.5 years, while for the former it is 2.75 years.}
\end{figure}

\begin{figure}[tbp]
\epsfxsize=14cm
\centering{\ \epsfbox{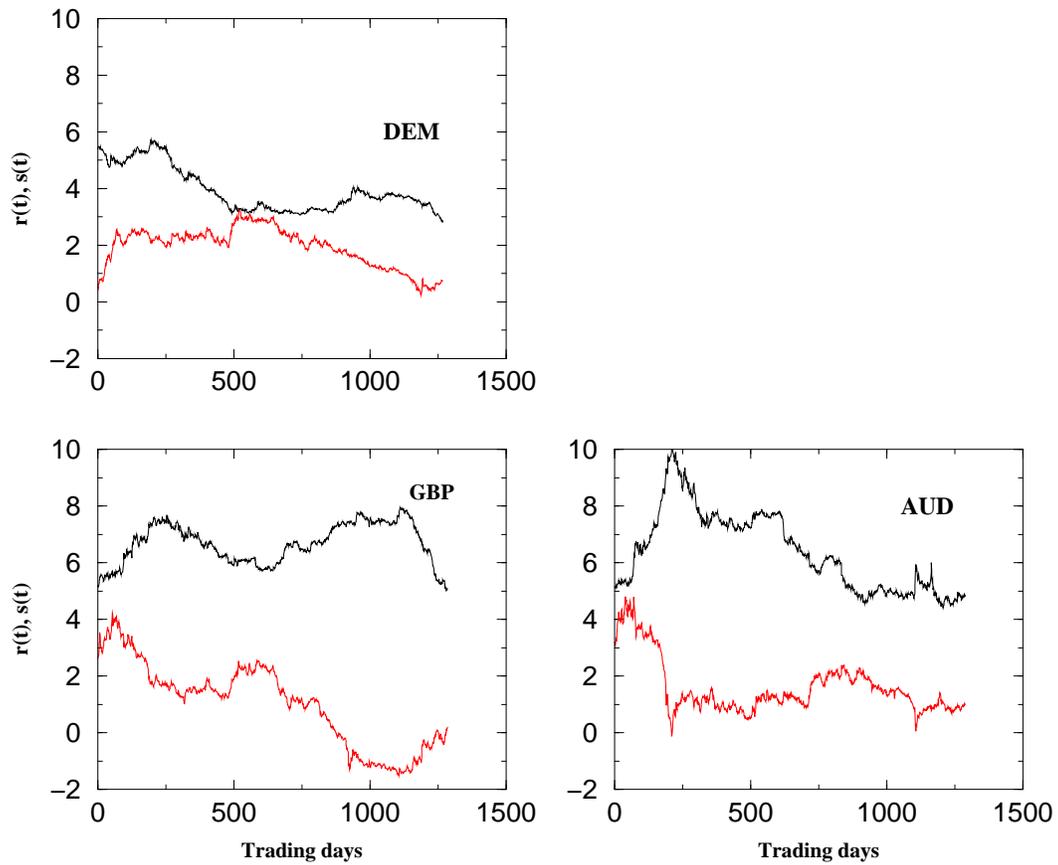}}
\vspace{0.0cm}
\caption{The historical times series of the spot rate 
(top bold curves) and long spread for the {\sc dem}, {\sc gbp} and {\sc aud}. 
For these cases the long spread is defined using the maximum forward rate 
of 2.75 years.}
\end{figure}

\begin{figure}[tbp]
\epsfxsize=12cm
\centering{\ \epsfbox{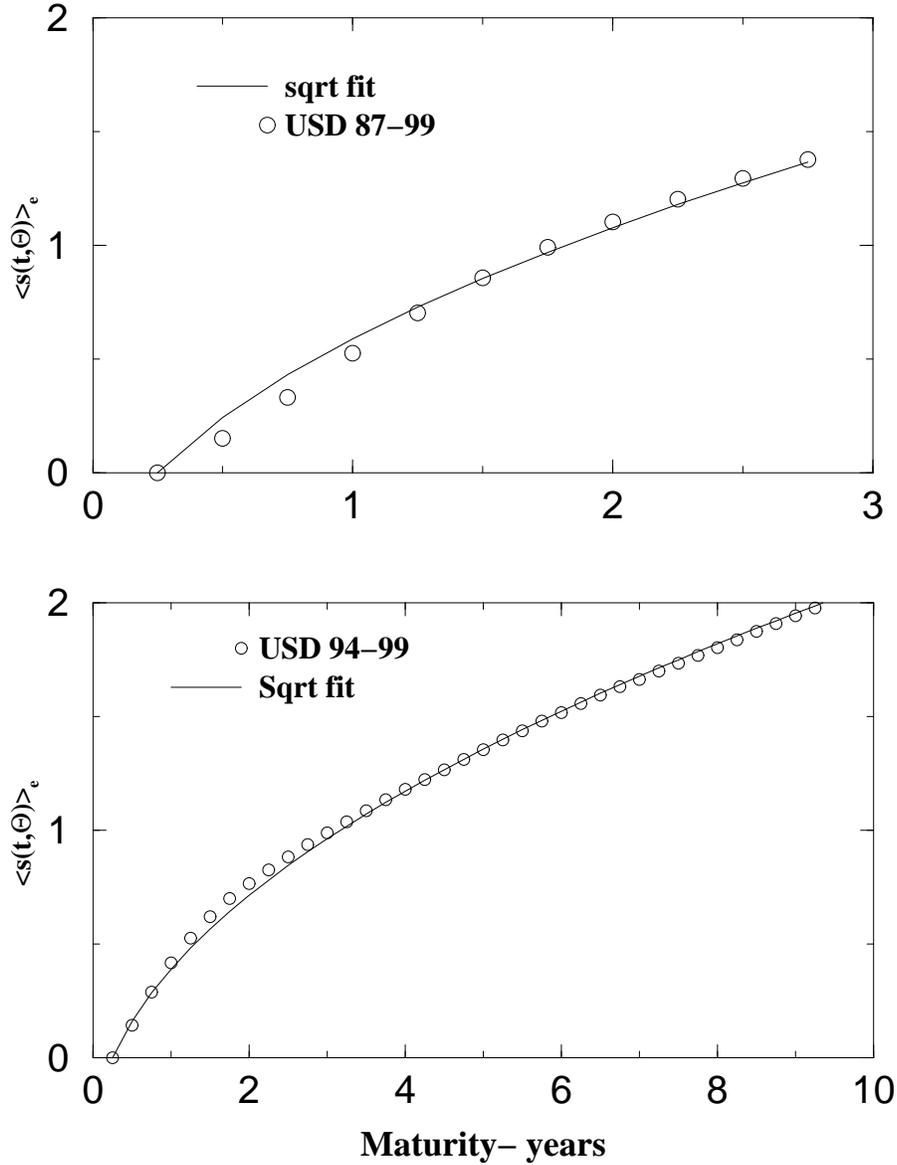}}
\vspace{0.0cm}
\caption{Top Figure: the average {\sc frc} in \% for 
{\sc usd} 87-99, given empirically by Eq. (3.1), and a best 
fit to Eq. (3.2). The fitting parameter $a$, is shown in Table 2.
Bottom Figure: the same but now for {\sc usd} 94-99. 
These figures, along with Table 2, demonstrate that the {\sc usd} 
average {\sc frc} is  
well fitted by a square-root law with a prefactor given approximately by the 
spot volatility.}
\end{figure}

\begin{figure}[tbp]
\epsfxsize=15cm
\centering{\ \epsfbox{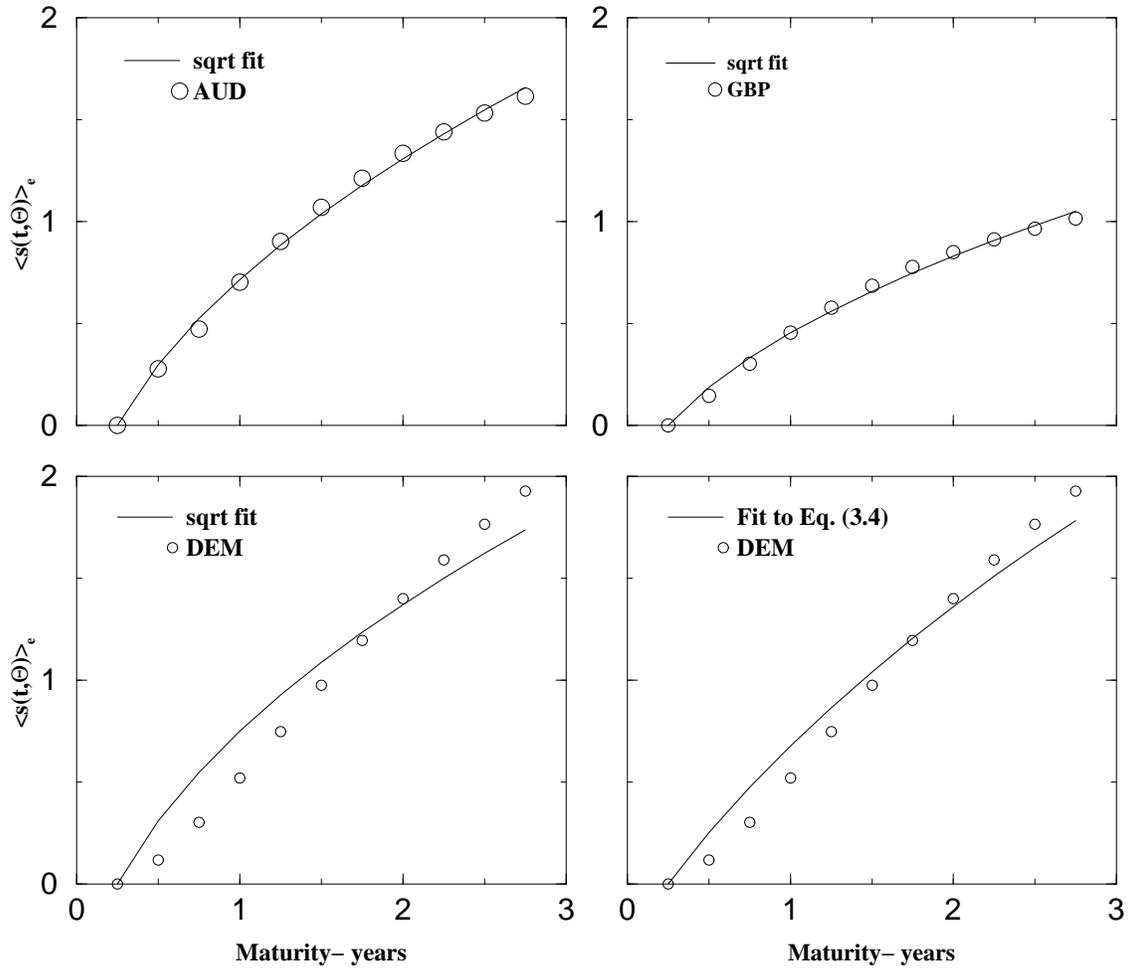}}
\vspace{0.0cm}
\caption{Top left: the same as figure 3 but now for {\sc aud}. Top right: 
the same but now for {\sc gbp}. 
Bottom left: the same but now for {\sc dem}. 
Bottom right: here we show the same empirical {\sc dem} average {\sc frc},
but now with a best fit to Eq. (3.4) rather than Eq. (3.2). 
These figures, along with Table 2, demonstrate that the {\sc aud} 
and {\sc gbp} average {\sc frc}'s
are also well fitted by a square root law with a prefactor given approximately 
by the spot volatility. In section 3.2 we explain why the {\sc dem} average {\sc frc} 
cannot be well fitted by a square-root law.}
\end{figure}

\begin{figure}[tbp]
\epsfxsize=15cm
\centering{\ \epsfbox{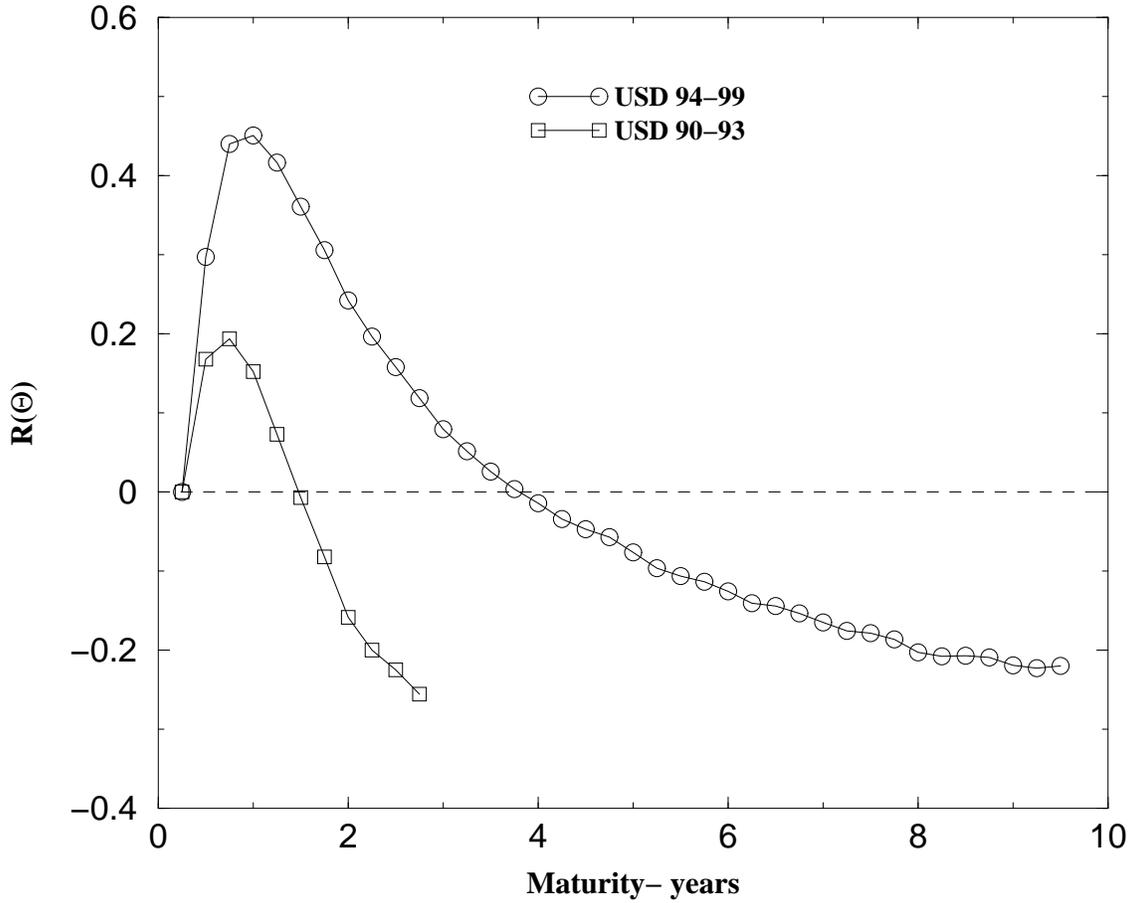}}
\vspace{0.0cm}
\caption{The {\sc usd} response function, derived empirically 
from Eq. (3.7). We show results for the dataset {\sc usd} 94-99
and the 4 year dataset {\sc usd} 90-93. This figure demonstrates 
a positive peak at a maturity of 0.75-1 year.}
\end{figure}

\begin{figure}[tbp]
\epsfxsize=15cm
\centering{\ \epsfbox{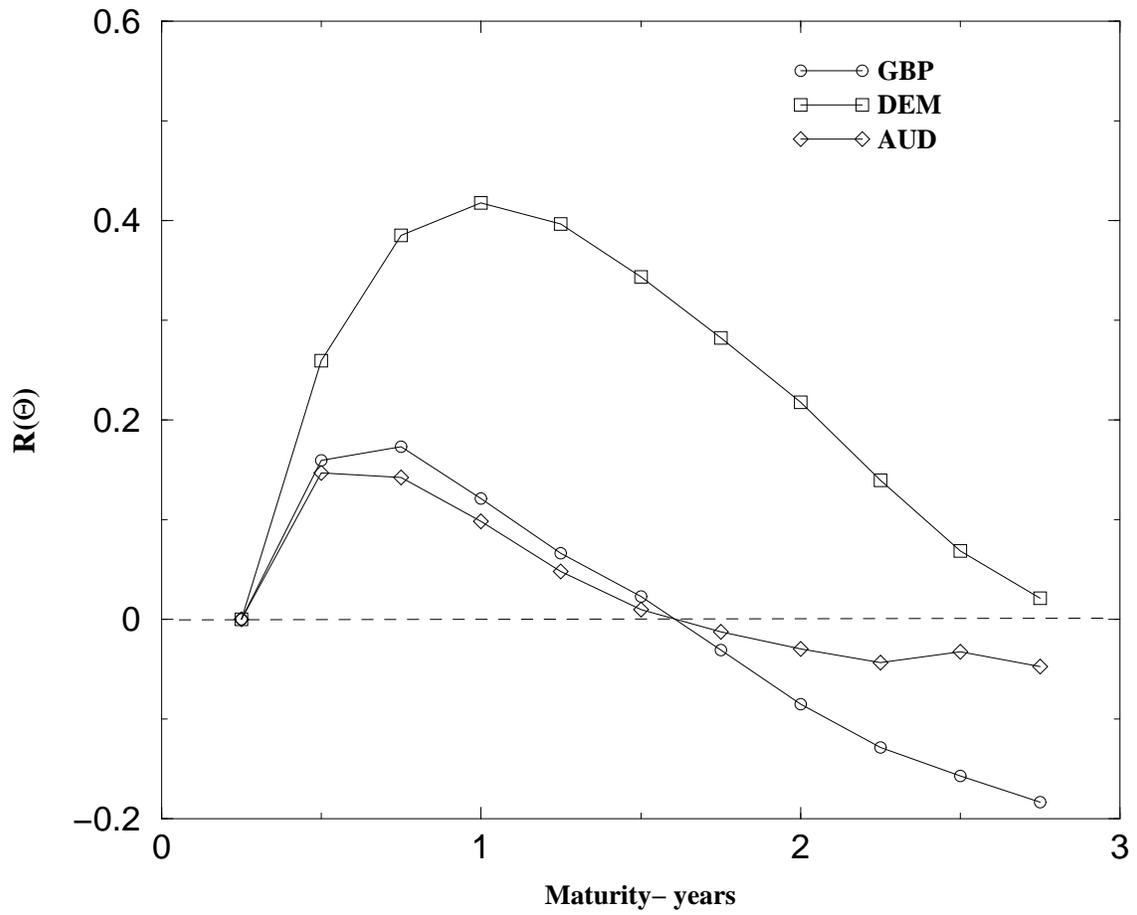}}
\vspace{0.0cm}
\caption{The same as figure 5 but now for {\sc gbp}, {\sc dem} and {\sc aud}.
This figure demonstrates that the positive peak found in figure 5 is robust 
across currencies as well as time periods.}
\end{figure}

\begin{figure}[tbp]
\epsfxsize=15cm
\centering{\ \epsfbox{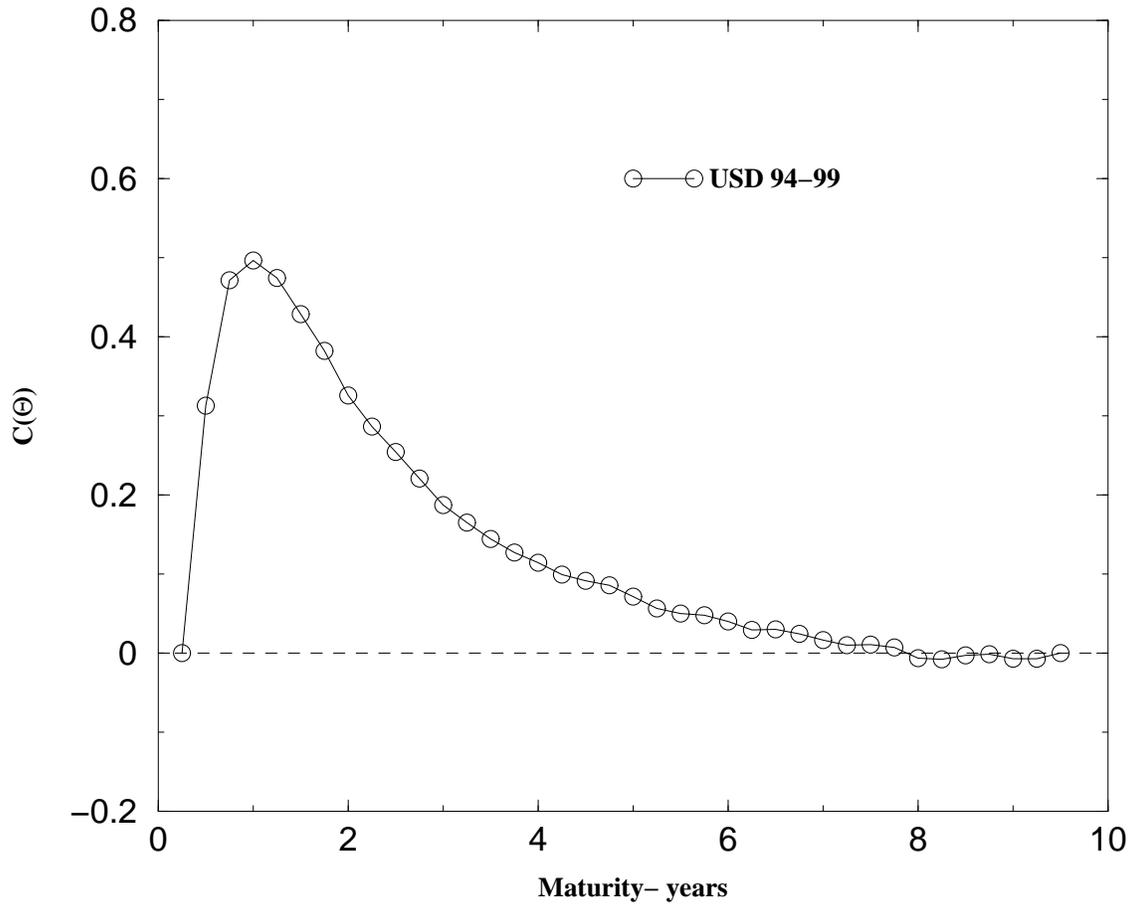}}
\vspace{0.0cm}
\caption{The anticipated trend scaling function, $C(\theta)$, defined in 
Eq. (4.11), for the {\sc usd} 94-99. This function is simply deduced from 
Eq. (4.18), 
using previous empirical results for the response function 
and the average {\sc frc}. The shape of this function is a consequence of the 
positive peak shown in figure 5 for the response function.}
\end{figure}

\begin{figure}[tbp]
\epsfxsize=15cm
\centering{\ \epsfbox{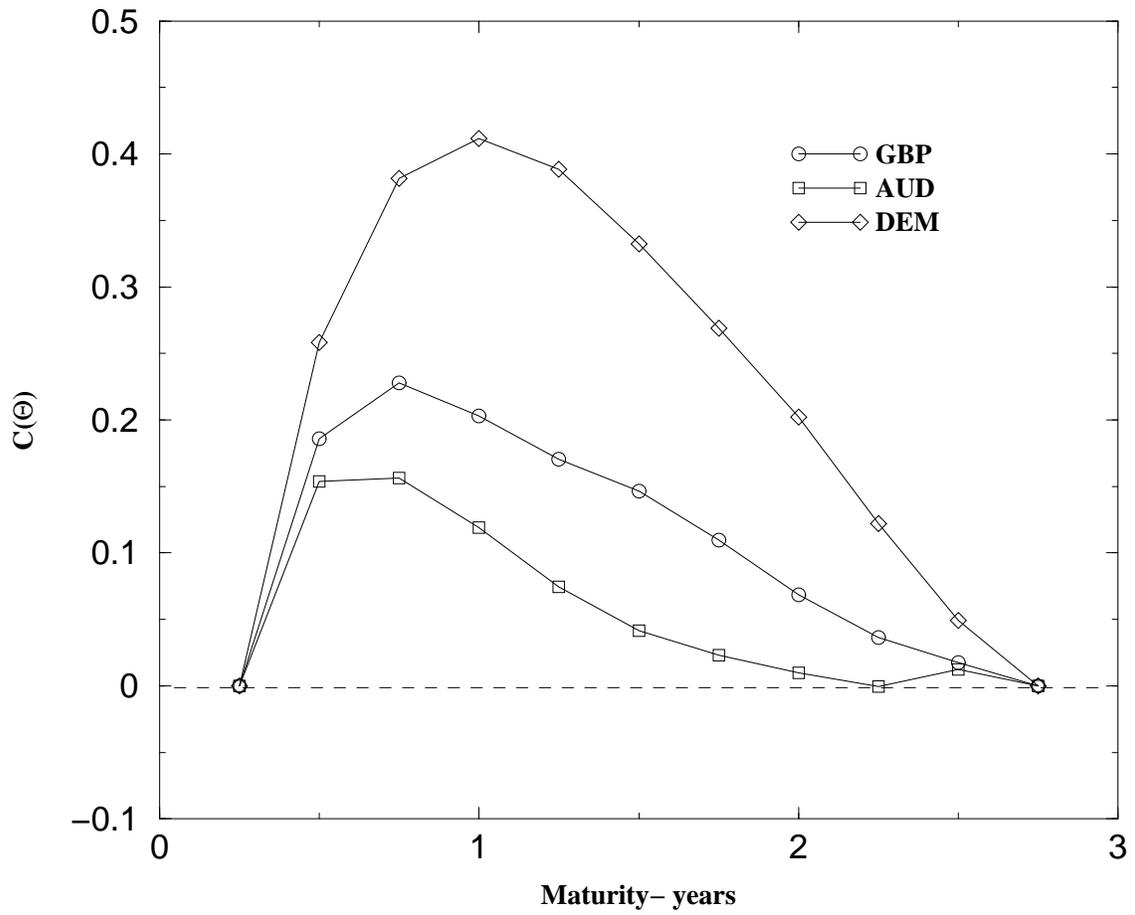}}
\vspace{0.0cm}
\caption{The same as figure 7 but now for {\sc gbp}, {\sc dem} and {\sc aud}. 
Again, the shape of these functions are a consequence of the positive peaks 
shown in figure 6.}
\end{figure}

\begin{figure}[tbp]
\epsfxsize=14cm
\centering{\ \epsfbox{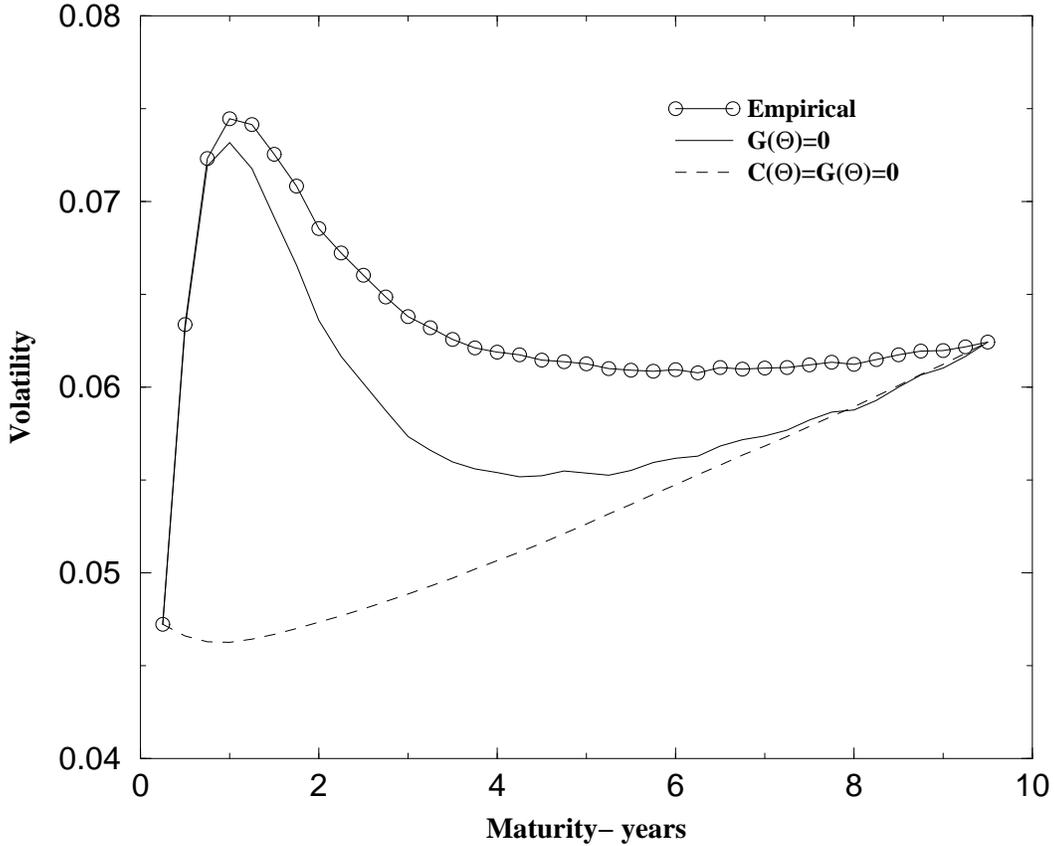}}
\vspace{0.0cm}
\caption{{\sc frc} volatility for {\sc usd} 94-99 in units of 
\% per square-root day. We show the empirical and model volatilities. The 
empirical volatility is given by Eq. (3.8).
The model volatilities are given by the square-root of Eq. (4.19) when $C=G=0$ and $G=0$.
We have used $Q_K=2$. In the former case the model has only been calibrated to 
the spot and long spread volatility 
and the average {\sc frc}. In the latter case the model is also 
calibrated to the response function. The figure demonstrates how the 
response function determines the qualitative shape of the {\sc frc} volatility.}
\end{figure}

\begin{figure}[tbp]
\epsfxsize=14cm
\centering{\ \epsfbox{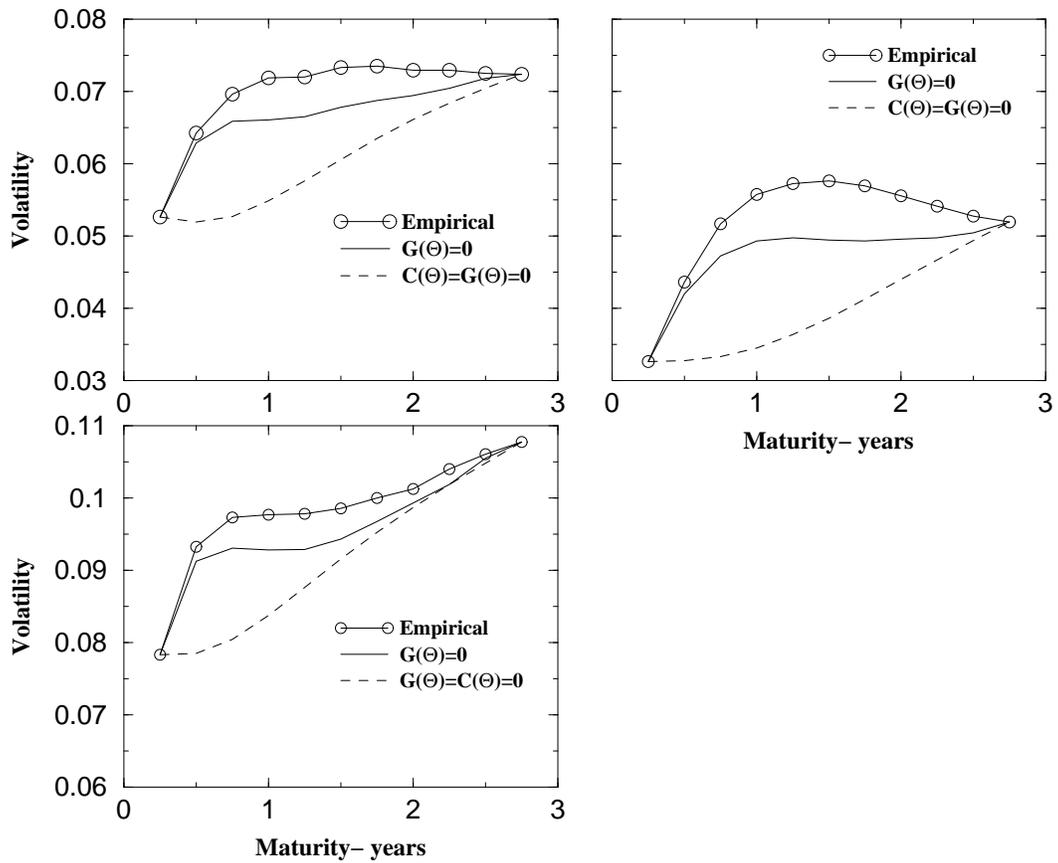}}
\vspace{0.0cm}
\caption{Same as for Figure 9 but now for {\sc gbp} (top left), 
{\sc dem} (top right) and {\sc aud}(bottom left). 
This figure demonstrates that for all the currencies, the response 
function determines the qualitative shape of the {\sc frc} volatility.}
\end{figure}

\begin{figure}[tbp]
\epsfxsize=12cm
\centering{\ \epsfbox{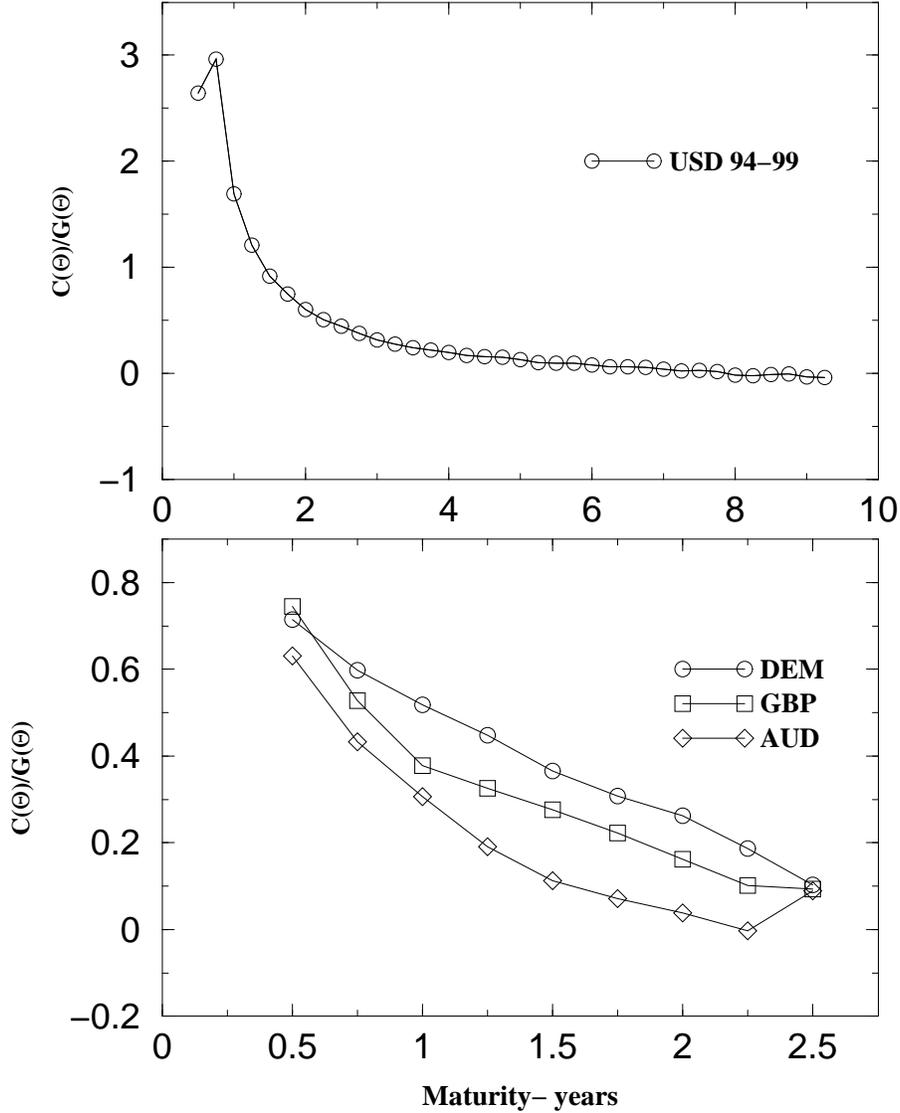}}
\vspace{0.0cm}
\caption{Top figure: The deformation signal to noise ratio: $C(\theta)/G(\theta)$ 
for the {\sc usd} 94-99. Bottom figure: the same but now for the currencies 
{\sc gbp}, {\sc dem} and {\sc aud}. 
First $C(\theta)$ is 
determined from Eq. (4.18), using the empirical average {\sc frc} and 
the response function. 
Then $G(\theta)$ is calibrated to the empirical {\sc frc} volatility using 
Eq. (4.19). From Eq. (4.11) we see that the signal to noise ratio provides a 
relative measure of the spot and 
noise contributions to a deformation increment. 
These figures demonstrate that the {\sc usd} has a much 
stronger signal to noise ratio than the other currencies.}
\end{figure}

\begin{figure}[tbp]
\epsfxsize=14cm
\centering{\ \epsfbox{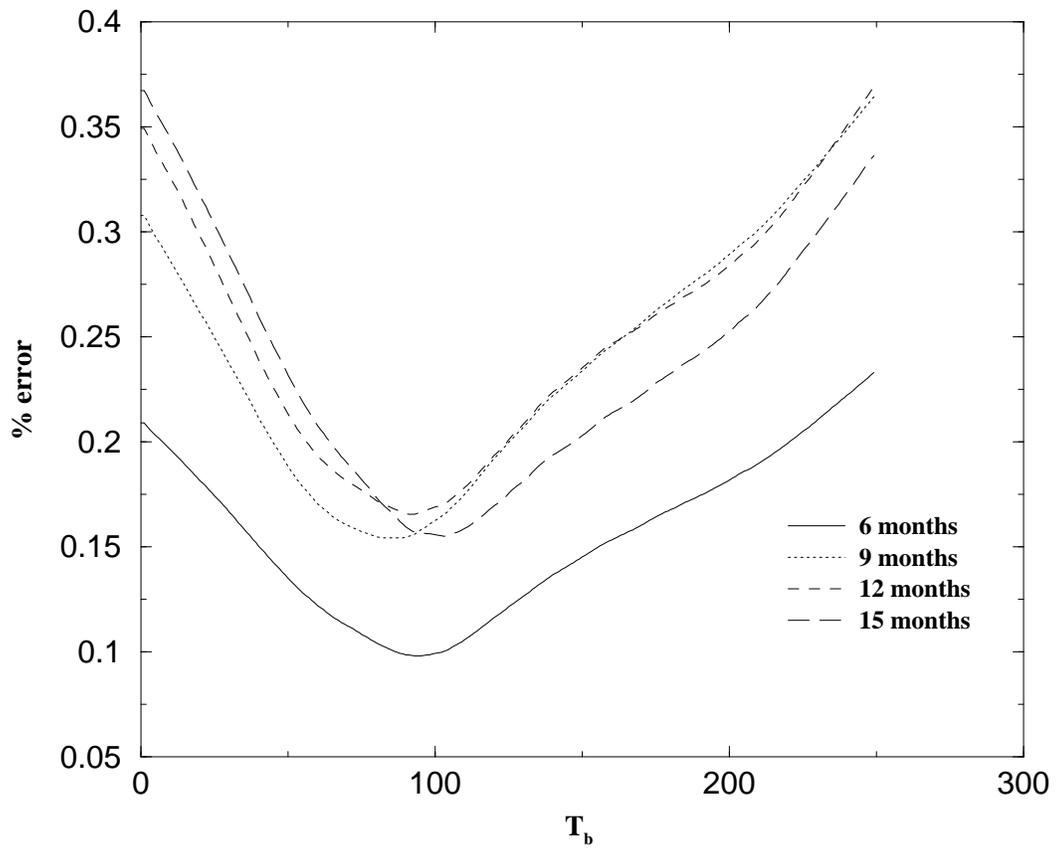}}
\vspace{0.0cm}
\caption{Plot the {\sc lhs} of Eq. (5.2), against the parameter 
$T_b$ (in trading days), where for 
the simulation of $b(t)$ we have used the {\sc fw} model, 
Eq. (4.16): for {\sc usd} 94-99.
This figure demonstrates the clear presence of a 100 trading 
day timescale in the {\sc usd} {\sc frc} dynamics.}
\end{figure}

\begin{figure}[tbp]
\epsfxsize=14cm
\centering{\ \epsfbox{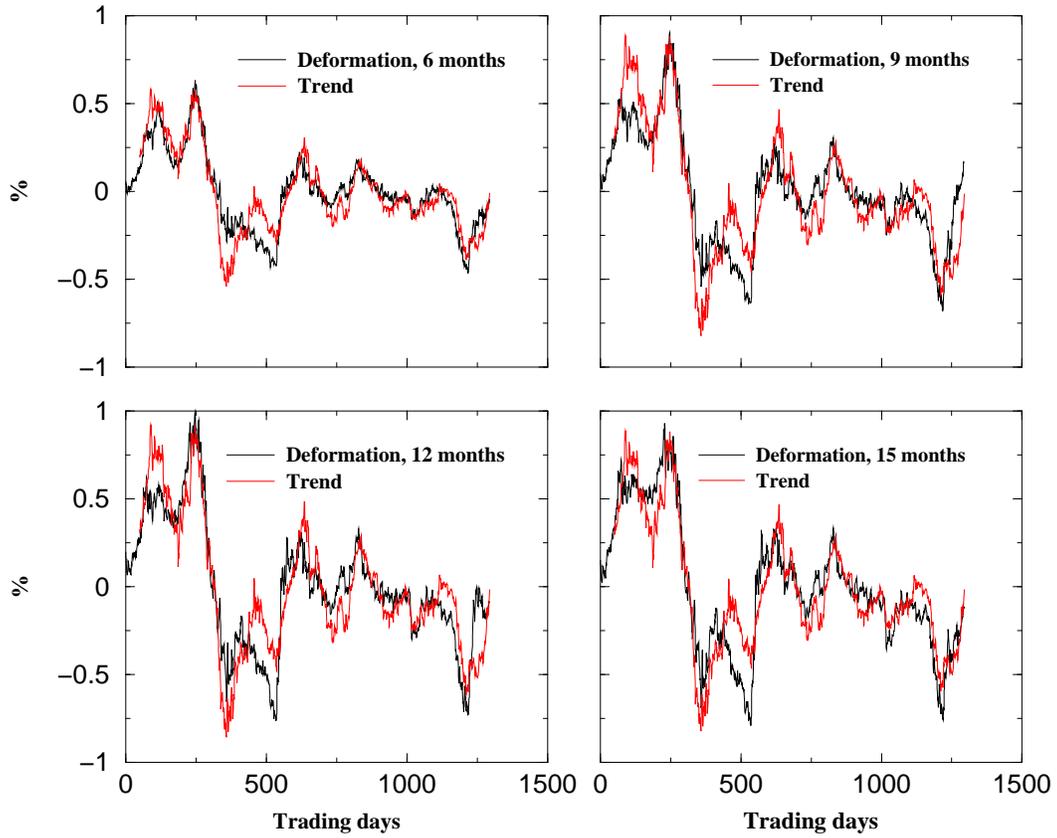}}
\vspace{0.0cm}
\caption{Comparison for {\sc usd} 94-99, of the empirical deformation process 
and the scaled anticipated trend, $C(\theta)b(t)$, where $b(t)$ is calculated 
using the {\sc fw} model, Eq. (4.16), with $T_b=100$ days as suggested by 
the minimum of figure 12. This figure 
demonstrates a striking correlation between the empirical deformation process 
and the anticipated trend.}
\end{figure}

\begin{figure}[tbp]
\epsfxsize=14cm
\centering{\ \epsfbox{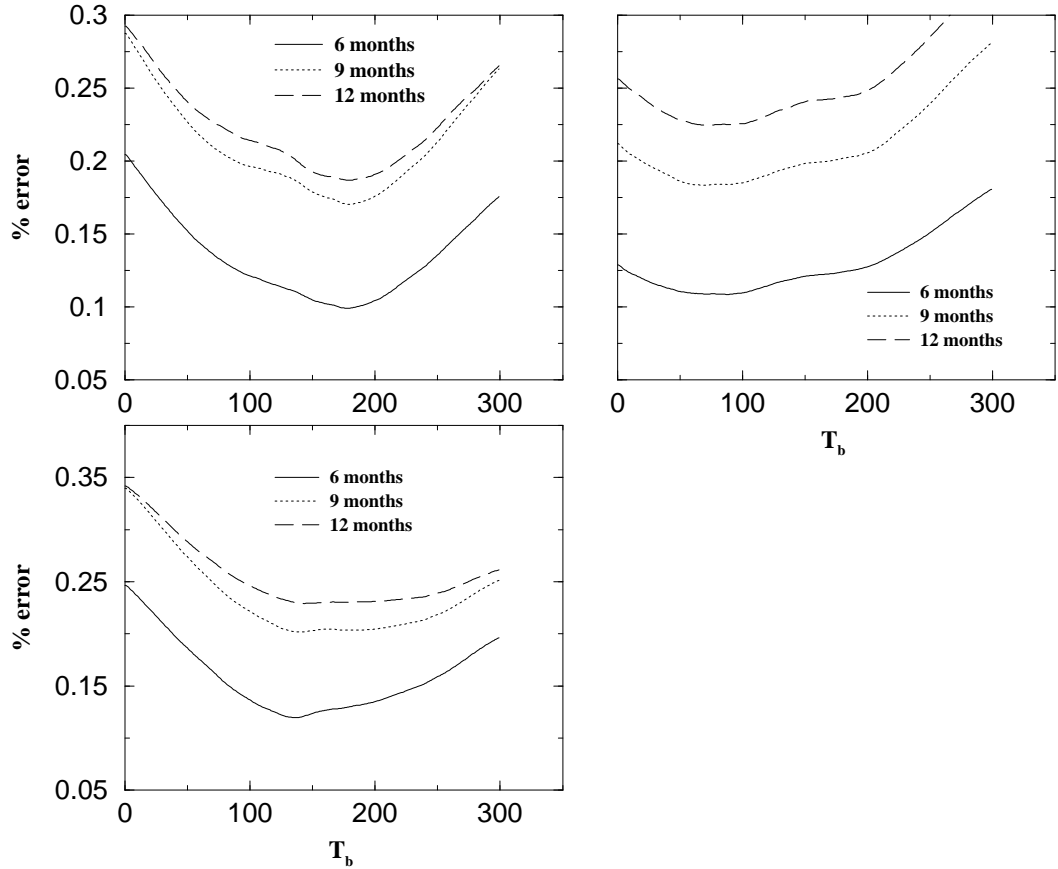}}
\vspace{0.0cm}
\caption{Top left: Same as for figure 12 but now for the {\sc gbp}. 
Top right: for the {\sc dem}. Bottom left:  for the {\sc aud}. These figures
demonstrates the same qualitative features as found in figure 12 for the {\sc usd}.
Here the minimums are not as well defined. This is expected due to the weaker signal 
to noise ratio shown in figure 11.}
\end{figure}

\begin{figure}[tbp]
\epsfxsize=14cm
\centering{\ \epsfbox{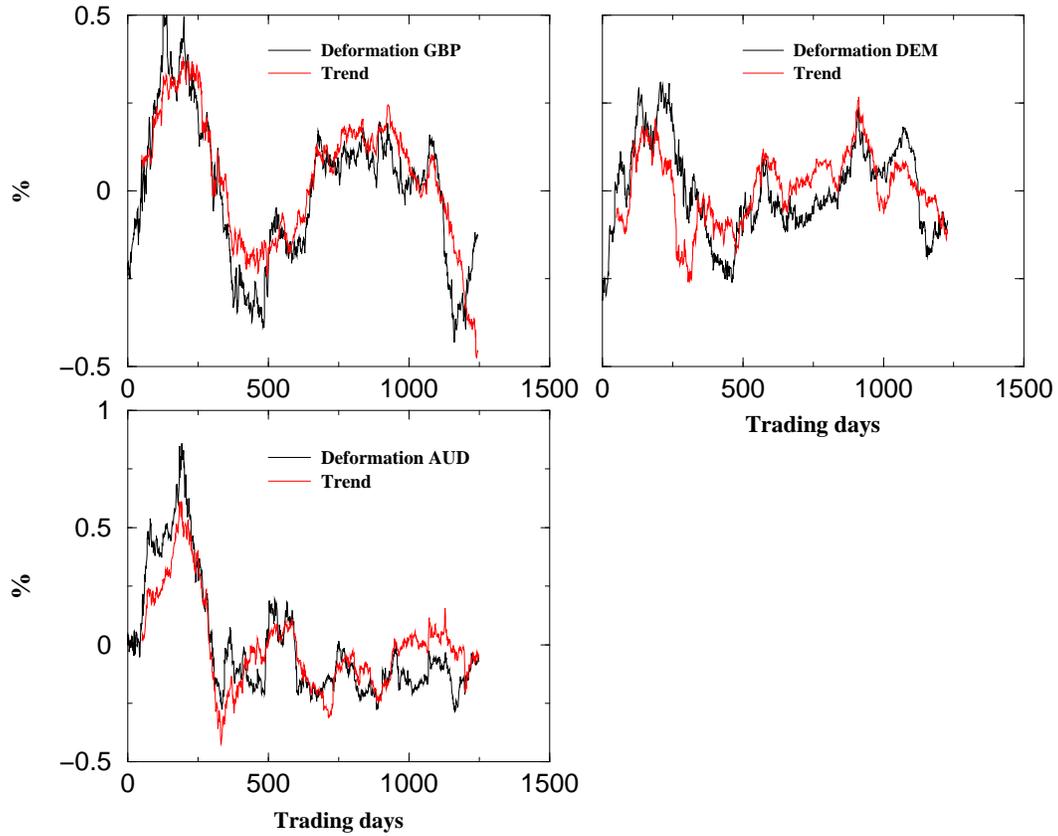}}
\vspace{0.0cm}
\caption{Top left: Comparison for {\sc gbp}, of the empirical deformation process 
and the scaled anticipated trend, $C(\theta)b(t)$, where $b(t)$ is calculated 
using the {\sc fw} model, Eq. (4.16), with $T_b=175$ days as suggested by 
minimum of figure 14. We have used the 6 month 
maturity. Top right: the same but now for the {\sc dem} with $T_b=100$ days. 
Bottom left: the same but now for the {\sc aud} with $T_b=130$ days. 
These figures demonstrate a clear correlation between the empirical deformation 
process and the anticipated trend. The correlation is not as striking as for 
the {\sc usd} due to the lower signal to noise ratio shown in figure 11.}
\end{figure}

\begin{figure}[tbp]
\epsfxsize=14cm
\centering{\ \epsfbox{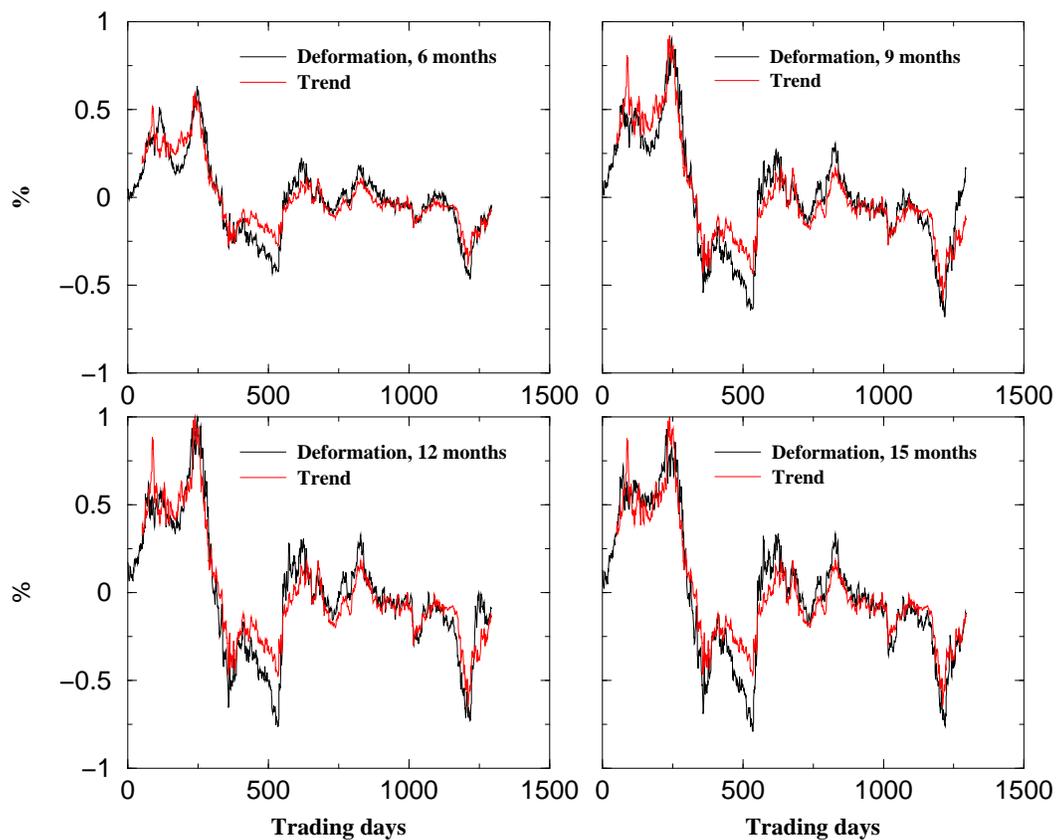}}
\vspace{0.0cm}
\caption{Comparison for {\sc usd} 94-99, of the empirical deformation process 
and the scaled anticipated trend, $C(\theta)b(t)$. In this case $b(t)$ is calculated 
using the {\sc ou} model, Eq. (4.15), with $\lambda_b^{-1}=100$ days, while $C(\theta)$ is 
calibrated to the {\sc frc} volatility, Eq. (4.19), where $G(\theta)$ has been set 
to zero (2 factor model). This figure demonstrates an especially
 spectacular correlation 
between the empirical deformation process and the anticipated trend.}
\end{figure}

\begin{figure}[tbp]
\epsfxsize=14cm
\centering{\ \epsfbox{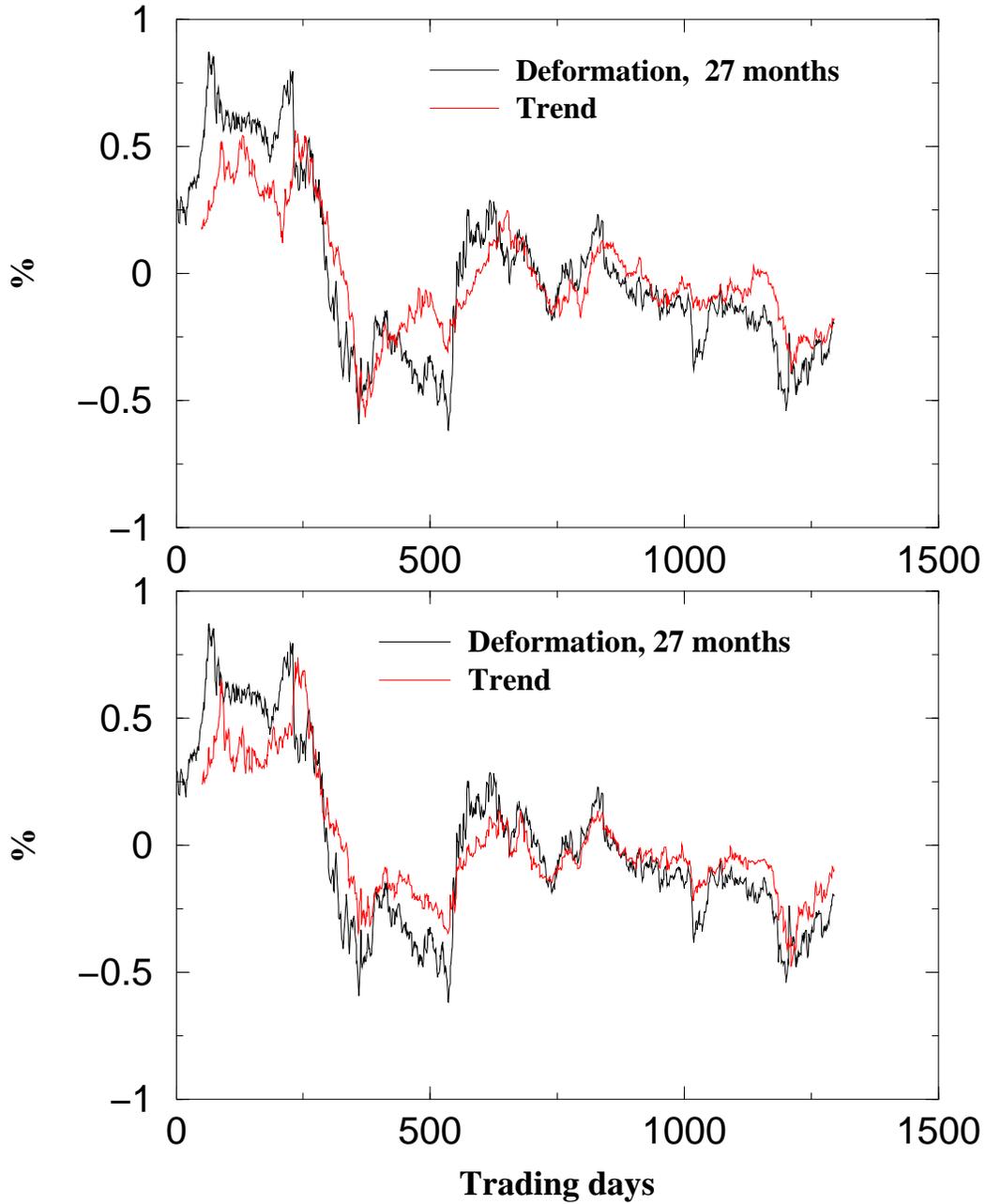}}
\vspace{0.0cm}
\caption{Top Figure: the same as for figure 13 but now for a maturity of 27 months.
Bottom Figure: the same as for figure 16 but also now for a maturity of 27 months. 
This figure demonstrates that the correlation between the empirical deformation 
and anticipated trend persists even 2 years forward of the spot.}
\end{figure}

\vfill
\eject

\end{document}